\documentclass[12pt,a4paper]{article}
\usepackage[pdftex,colorlinks=true]{hyperref}
\usepackage{amsmath,caption,setspace,dsfont,xcolor,orcidlink,multirow}
\definecolor{darkblue}{rgb}{0,0,.6}
\hypersetup{citecolor=darkblue,linkcolor=darkblue,urlcolor=darkblue}
\usepackage[top=1.15in, bottom=1.15in, left=1.1in, right=1.1in]{geometry}

\DeclareCaptionStyle{italic}[justification=centering]
 {labelfont={bf},textfont={it},labelsep=colon}
\usepackage{graphicx,psfrag,epsf}
\usepackage{enumerate}
\usepackage{natbib}
\usepackage{url}
\usepackage{booktabs, subfig, bm, paralist,mathpazo,tikz,todonotes,longtable,microtype}
\newcommand{\Rlogo}{\protect\includegraphics[height=1.8ex,keepaspectratio]{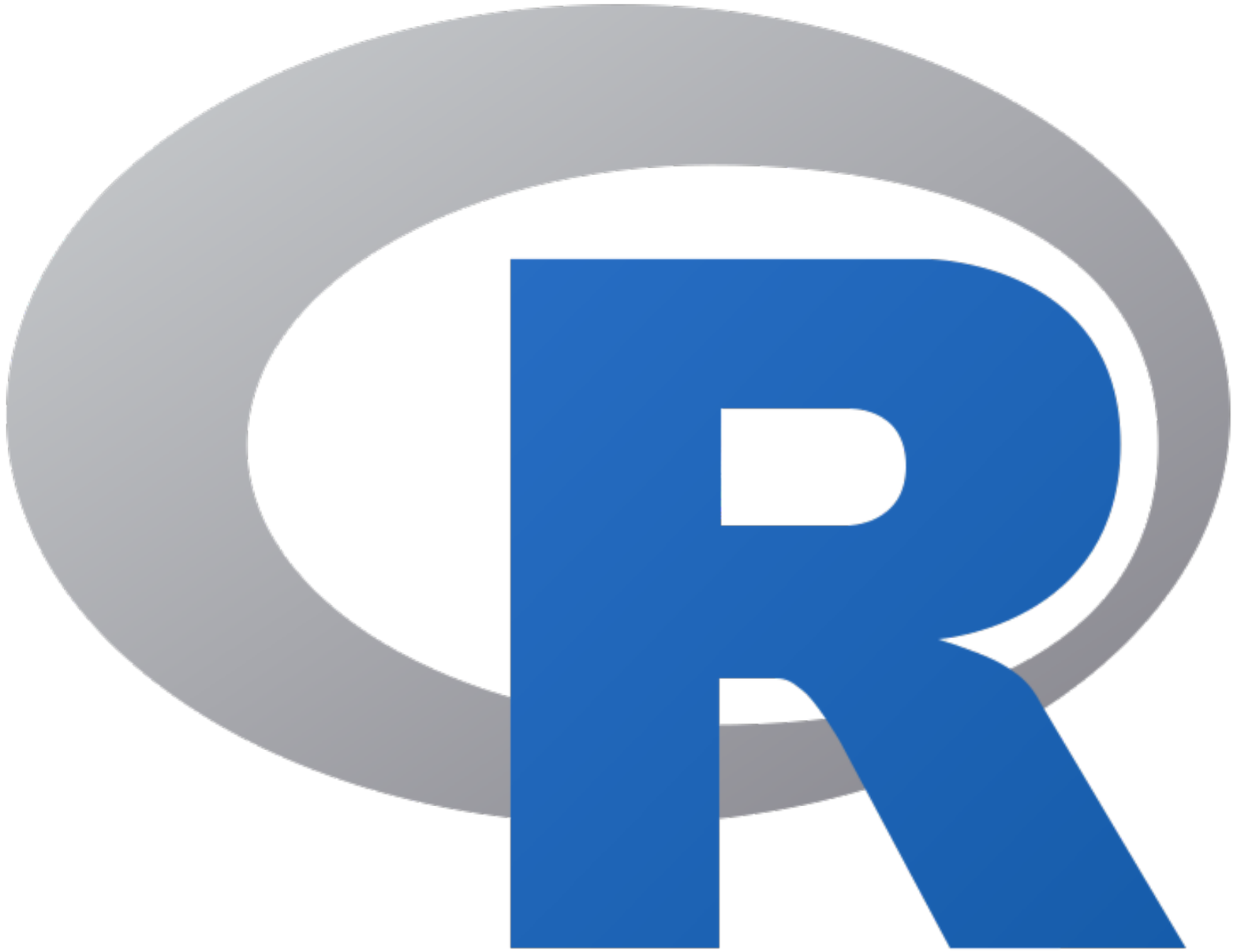}}

\newcommand{\blind}{0}

\newcommand{\X}{\mathcal{X}}
\newcommand{\Y}{\mathcal{Y}}
\newcommand{\Z}{\mathcal{Z}}
\graphicspath{{plots/}}
\DeclareMathOperator*{\argmin}{\arg\!\min}
\newsavebox\CBox
\def\textBF#1{\sbox\CBox{#1}\resizebox{\wd\CBox}{\ht\CBox}{\textbf{#1}}}

\newcommand\omicron{o}

\definecolor{a0}{rgb}{0.0, 0.5, 0.0}
\definecolor{bistre}{rgb}{0.24, 0.17, 0.12}
\definecolor{amethyst}{rgb}{0.6, 0.4, 0.8}
\definecolor{blue-violet}{rgb}{0.54, 0.17, 0.89}
\definecolor{Rcolor}{RGB}{150,160,190}
\definecolor{blush}{rgb}{0.87, 0.36, 0.51}
\definecolor{brightturquoise}{rgb}{0.03, 0.91, 0.87}
\definecolor{burntorange}{rgb}{0.8, 0.33, 0.0}
\usepackage{array}
\newcolumntype{L}[1]{>{\raggedright\arraybackslash}p{#1}}

\begin{document}

\def\spacingset#1{\renewcommand{\baselinestretch}
{#1}\small\normalsize} \spacingset{1}

\if0\blind
{
\title{\bf Forecasting multiple density-valued time series}
\author{\normalsize Han Lin Shang \orcidlink{0000-0003-1769-6430}\thanks{Corresponding author: Department of Actuarial Studies and Business Analytics, Macquarie University, 4 Eastern Road, Sydney NSW 2109, Australia. Telephone: +61(2) 9850 4689; E-mail: hanlin.shang@mq.edu.au.}\hspace{.2cm}\\
\normalsize  Department of Actuarial Studies and Business Analytics \\
\normalsize  Macquarie University}
\date{}
\maketitle
} \fi

\if1\blind
{
\bigskip
\bigskip
\bigskip
\begin{center}
{\LARGE\bf Forecasting multiple density-valued time series}
\end{center}
  \medskip
} \fi

\bigskip
\begin{abstract}
Density-valued time series are common in practice. However, since these functions are non-negative and must integrate to one, they do not form a linear vector space, which complicates the direct application of standard time-series methods. To address these constraints, we introduce two transformations, based on the centered log-ratio transformation and the cumulative distribution function, for modeling and forecasting balanced panel data represented by multiple densities. Both transformations map densities bijectively into an unconstrained function space in which cross-sectional and serial dependence can be modeled directly. Within the resulting unconstrained space, we consider two approaches: multi-population functional time-series modeling to account for correlations among populations, and gap modeling between gender and region. Using Japanese subnational age-specific life-table death counts from 1973 to 2024, we evaluate and compare the one- to 20-step-ahead point and interval forecast accuracy of the two approaches for each transformation, and offer some general recommendations. \\

\noindent \textit{Keywords}: centered log-ratio transformation; cumulative distribution function transformation; constrained time series; density function forecasting; Japanese subnational age distribution of deaths
\end{abstract}

\newpage
\spacingset{1.48} 

\section{Introduction}\label{sec:intro}

Multiple correlated density-valued time series are often observed simultaneously, with examples including income distributions across different populations \citep{KU01}, financial return distributions for multiple stocks \citep{PZK22}, distributions of bidding times in online auctions for various items \citep{WJS08}, age distribution of fertility across countries \citep{MS15}, and subnational age distributions of deaths in demography \citep{JSS25}, among others. 

We study modeling and forecasting age distribution of deaths, also known as life-table death counts, at subnational levels. Life-table death counts of a prefecture over time form one density-valued time series, while geographically neighboring prefectures generate a collection of correlated density-valued time series. The predicted subnational life-table death counts are useful for applications in actuarial science, such as annuity pricing and reserve planning.

Modeling these series separately ignores their cross-sectional dependence and may fail to capture the dominant common dynamics shared across prefectures. In response to this problem, we approach it through multi-population modeling \citep{EKC17} or gap modeling \citep{PCV18}. In the former approach, the essence is to represent a multidimensional process through a small number of latent factors, thereby capturing the dominant common dynamics in a lower-dimensional space. Some common techniques include functional analysis of variance \citep[see, e.g.,][]{Zhang14} and functional factor models \citep[see, e.g.,][]{GQW25, LLS+26}. In the latter approach, we explore the gap between gender and region; by modeling and forecasting these gaps, we use high-quality data, such as female or national data, to improve the forecasts of relatively low-quality data, such as male or subnational data \citep[see, e.g.,][]{SN26}.

Building on a \textit{one-to-one} mapping, we introduce two transformations that map the nonlinear, constrained space to a linear, unconstrained space. The two transformations are centered log-ratio (clr) transformation \citep[see, e.g.,][]{Aitchison86} and cumulative distribution function (CDF) transformation \citep[see, e.g.,][]{SH25}. The clr transformation is the isometric isomorphism that maps Bayes Hilbert space onto $L^2$ function space \citep{MSF+25}. The CDF transformation handles the presence of zero values and imposes monotonicity. Both transformations present a measure, in which standard algebraic operations, such as addition and scalar multiplication, are possible.

The contributions of this paper are twofold.
\begin{inparaenum}
\item[First,] we revisit some multi-population modeling and forecasting methods, and introduce a novel combination of two existing tools, based on two-way functional analysis of variance and functional factor model, under the clr and CDF transformations.
\item[Second,] between the multi-population modeling and gap modeling, we evaluate and compare their point and interval forecast accuracies. Such a comparative study has not been explored previously, and it may help identify the best-performing forecasting method under each transformation.
\end{inparaenum}

The paper is structured as follows: In Section~\ref{sec:2}, we describe Japanese subnational life-table death count data set. In Section~\ref{sec:3}, we review the clr and CDF transformations. In Section~\ref{sec:4}, we consider some multi-population models to forecast multiple unconstrained functional time series under each transformation. In Section~\ref{sec:5}, we describe the gap forecasting methods, particularly the gender gap, region gap and double gap. Section~\ref{sec:6} describe a split conformal prediction for constructing pointwise prediction intervals. Section~\ref{sec:7} evaluates and compares point forecast accuracy using the Kullback-Leibler divergence (KLD) in~\eqref{eq:KLD} and interval forecast accuracy using the coverage probability difference and mean interval score. In Section~\ref{sec:8}, we conclude with some ideas on how the methodology presented here can be further extended.

\section{Japanese subnational age distribution of deaths}\label{sec:2}

Understanding patterns in mortality across subpopulations is essential for local health policy decision-making. This insight has led to a recent line of work focused on studying disparities in mortality outcomes within countries at the subnational level. In the U.S., researchers are analyzing differences in life expectancy across states \citep[see, e.g.,][]{WS19, HRK21}. In recent years, it has become increasingly clear that disparities in mortality outcomes along demographic dimensions such as sex, geography, socioeconomic status, and education are growing \citep[see, e.g.,][]{Hendi15, KEB+20}. In response, estimates of subpopulation mortality risks are required to identify and understand the mortality patterns of vulnerable groups, track the effects of policy response, and allocate current and future resources.

Sourced from \cite{JMD26}, we consider Japanese period life-table death counts by age, sex, and region. For a given calendar year $t$ from 1973 to 2024, we observe life-table death counts, denoted by $d_{t,x}^{s,g}$ for region~$s$, gender~$g$, and age~$x$; it is defined as the number of deaths occurring between the two successive ages in a period life table. By construction, life-table death counts are non-negative and sum to a radix of $10^5$ for each year. After normalization (dividing by its radix), it resembles a probability density function (PDF).

\begin{figure}[!htb]
\centering
{\includegraphics[width=8.4cm]{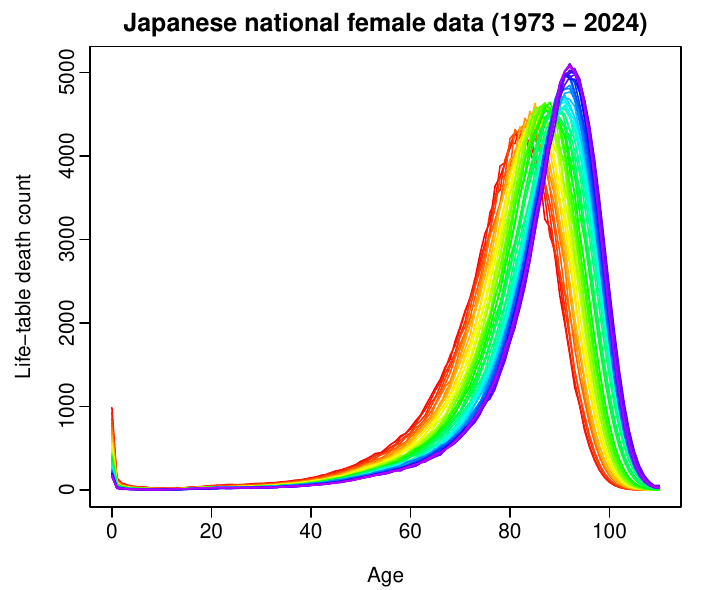}}
\quad
{\includegraphics[width=8.4cm]{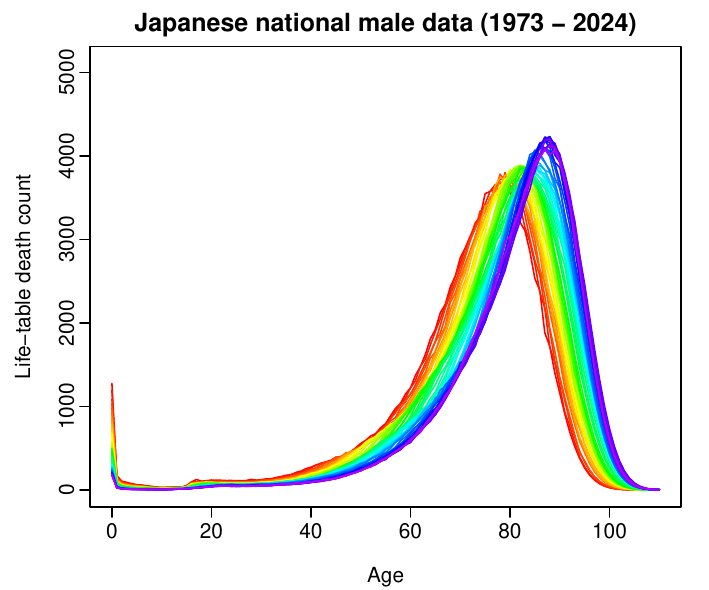}}
\caption{\small Functional time series graphical displays of Japanese age-specific life-table death counts from 1973 to 2024 in a single-year group. The life-table death counts are nonnegative and sum to a radix of $10^5$ for each year. The oldest years are shown in red, with the most recent years in violet. Curves are ordered chronologically according to the colors of the rainbow.}\label{fig:1}
\end{figure}

As shown in Figure~\ref{fig:1}, the time ordering of the curves follows the color order of a rainbow, where curves from the distant past are shown in red, and the more recent curves are shown in violet. The figures show typical mortality curves for mainly developed countries, with a decreasing trend in infant death counts due to modern medicine. A typical negatively skewed distribution for the life-table death counts is apparent, with peaks shifting to higher ages for both females and males. This gradual shift is a key driver of longevity risk, which is a major issue for insurers and pension funds, especially in the selling and risk management of annuity products \citep[see][for a discussion]{DDG07}.

The rainbow plots allow visualization of a series at the national or subnational level; the image plot of KLD between subnational and national life-table death counts allows visualization of regional heterogeneity, particularly for prefectures with higher and lower mortality \citep[see, e.g.,][]{SH17, SN26}.

\section{A one-to-one mapping to unconstrained space}\label{sec:3}

\subsection{Centered log-ratio transformation}\label{sec:3.1}

For a given prefecture and gender in year $t$, life-table death counts can be defined as a random vector of $M=111$ positive components $(d^{s,g}_{t,1}, d^{s,g}_{t,2},\dots,d^{s,g}_{t,M})$ with non-negative values whose sum is a given constant, set typically equal to 1 (portions), 100 (\%), and $10^6$ parts per million (ppm) in geochemical trace element compositions \citep{Aitchison86}. The compositional data reside in a simplex
\begin{equation*}
\mathcal{S}^{M-1} = \left\{\bm{Y}_{t}^{s,g} = \Big(d_{t,1}^{s,g}, d_{t,2}^{s,g},\dots,d_{t,111}^{s,g}\Big)^{\top}, \quad d_{t,x}^{s,g}\geq 0, \quad \sum^{M}_{x=1}d_{t,x}^{s,g} = c\right\},\qquad t=1, 2,\dots,T,
\end{equation*}
where $\mathcal{S}$ denotes a simplex, $c$ is a fixed constant, $^{\top}$ denotes vector transpose and the simplex sample space is a $(M-1)$-dimensional subset of real-valued space $\mathbb{R}^{M-1}$.

For analyzing compositional data, a common transformation is the clr transformation \citep[see, e.g.,][with demographic applications]{BSO+18}. The clr transformation can be summarized as follows:
\begin{enumerate}[1)]
\item \textit{Input data}. We begin from a data matrix $\bm{Y}^{s,g}$ of dimension $(T\times M)$ of the life-table death counts $d_{t,x}^{s,g}$ with $T$ rows representing the number of years and $M$ columns representing the number of~ages.
\item \textit{clr transformation}. We apply the clr transformation, given by
\begin{equation*}
z_{t,x}^{s,g} = \ln \left(\frac{d_{t,x}^{s,g}}{\theta_t^{s,g}}\right), 
\end{equation*}
where $\theta_{t}^{s,g}$ denotes the geometric mean over the age at time $t$, given by
\begin{equation*}
\theta_t^{s,g} = \exp^{\frac{1}{M}\sum^M_{x=1}d_{t,x}^{s,g}}.
\end{equation*}
The clr-transformed data $\bm{z}_t^{s,g} = (z_{t,1}^{s,g},\dots,z_{t,M}^{s,g})$ reside in real-valued space.
\item \textit{Principal component analysis.} Principal component analysis is used to reduce dimensionality by summarizing the main mode of variations with only a few leading orthonormal principal components and their associated principal component scores. For a given prefecture and gender, we obtain
\begin{equation}
z_{t,x}^{s,g} = \widehat{\mu}_x^{s,g} + \sum^{L^{s,g}}_{l=1}\widehat{\beta}_{t,l}^{s,g}\widehat{\phi}_{l,x}^{s,g} + w_{t,x}^{s,g},\label{eq:PCA}
\end{equation}
where $\widehat{\mu}_x^{s,g}=\frac{1}{T}\sum^{T}_{t=1}z_{t,x}^{s,g}$ denotes the estimated mean term for age $x$ in year $t$, $w_{t,x}^{s,g}$ denotes model residual term, $\bm{\widehat{\Phi}}_x^{s,g} = \{\widehat{\phi}_{1,x}^{s,g},\widehat{\phi}_{2,x}^{s,g},\dots,\widehat{\phi}_{L^{s,g},x}^{s,g}\}$ represents the first $L^{s,g}$ sets of estimated principal components. To determine the number of components $L^{s,g}$, we consider an eigenvalue ratio (EVR) criterion of \cite{LRS20}. The value of $L^{s,g}$ is determined as the integer minimizing the ratio of two adjacent empirical eigenvalues, given by 
\begin{equation}
L^{s,g} = \argmin_{1\leq \kappa\leq T}\left\{\frac{\lambda^{s,g}_{\kappa+1}}{\lambda^{s,g}_{\kappa}}\times \mathds{1}\{\lambda^{s,g}_{\kappa}>\mathrm{r}\} + \mathds{1}\{\lambda^{s,g}_{\kappa}<\mathrm{r}\} \right\}, \label{eq:EVR}
\end{equation}
where $\lambda^{s,g}_{\kappa}$ is the $\kappa$\textsuperscript{th} empirical eigenvalue, computed via eigen-decomposition of the sample covariance of $\bm{z}^{s,g}$, $\mathds{1}\{\cdot\}$ represents binary indicator function, and customarily a threshold parameter $\mathrm{r} = 10^{-3}$ \citep[see, e.g.,][]{Shang26}.
\item \textit{Forecast of principal component scores.} Via a univariate or multivariate time-series forecasting method, such as exponential smoothing (ETS), we obtain the $h$-step-ahead forecast of the $l$\textsuperscript{th} principal component score $\widehat{\beta}^{s,g}_{T+h|T,l}$, where $h$ denotes the forecast horizon \citep[see also][]{HU07}. To select the optimal orders of the ETS, we use the corrected Akaike information criterion \citep{HT93}. Conditioning on the estimated principal components $\widehat{\bm{\Phi}}_x^{s,g}$, estimated mean term $\widehat{\mu}_x^{s,g}$, and observed data $\bm{z}_x^{s,g}=(z_{1,x}^{s,g}, z_{2,x}^{s,g},\dots,z_{T,x}^{s,g})$, the point forecast of $z_{T+h,x}^{s,g}$ can be obtained by
\begin{equation}
\widehat{z}_{T+h|T,x}^{s,g} = \text{E}\left[z_{T+h,x}^{s,g}|\bm{\widehat{\Phi}}_x^{s,g}, \widehat{\mu}_x^{s,g}, \bm{z}_x^{s,g}\right] = \widehat{\mu}_x^{s,g} + \sum^L_{l=1}\widehat{\beta}_{T+h|T,l}^{s,g}\widehat{\phi}_{l,x}^{s,g}.\label{eq:2}
\end{equation}
\item \textit{Transform back to the compositional data.} We take the inverse clr transformation, given by
\begin{equation*}
\widehat{\bm{d}}_{T+h}^{s,g} = \Bigg[\frac{\exp^{\widehat{z}_{T+h|T,1}^{s,g}}}{\sum^{M}_{x=1}\exp^{\widehat{z}_{T+h|T,x}^{s,g}}}, \frac{\exp^{\widehat{z}_{T+h|T,2}^{s,g}}}{\sum^{M}_{x=1}\exp^{\widehat{z}_{T+h|T,x}^{s,g}}}, \dots, \frac{\exp^{\widehat{z}_{T+h|T,M}^{s,g}}}{\sum^{M}_{x=1}\exp^{\widehat{z}_{T+h|T,x}^{s,g}}} \Bigg] \times 10^5,
\end{equation*}
where $\widehat{z}_{T+h|T,x}^{s,g}$ denotes the time-series forecasts in~\eqref{eq:2}, and $10^5$ is the life-table radix.
\end{enumerate}

In the clr transformation, we ought to take care of zero counts by Bayesian multiplicative replacement \citep[see, e.g.,][]{MHT+15, PM15}. Let $\varsigma$ be a detection limit, the multiplicative replacement of one-year life-table death count $d_{t,x}^{s,g}$ with $T_{0}$ zero cells is
\begin{equation} 
\check{d}_{t,x}^{s,g} = \left\{ \begin{array}{ll}
\varsigma & \mbox{if $d_{t,x}^{s,g} = 0$};\\
d^{s,g}_{t,x}(1-T_0\cdot \varsigma) & \mbox{if $d_{t,x}^{s,g} > 0$}.\end{array} \right. \label{eq:zero_replacement}
\end{equation}        
Commonly, $\varsigma = 0.65 \times \min_{x: d_{t,x}^{s,g}>0}d_{t,x}^{s,g}$ \citep[see, e.g.,][]{MHT+15}. Computationally, we implement the \textit{cmultRepl} function in the zCompositions package \citep{PM15}. Generally, we find that the results were not sensitive to the choice of $\varsigma$.

\subsection{Cumulative distribution function transformation}\label{sec:3.2}

We first normalize the life-table death counts by dividing each value by its radix, and let us denote $\widetilde{d}_{t,x}^{s,g}$ as the normalized life-table death counts. Via the cumulative sum, we transform a PDF into a CDF,
\begin{equation*}
D_{t,x}^{s,g} = \sum^x_{\omega=1}\widetilde{d}_{t,\omega}^{s,g},\qquad x=1, 2,\dots,111,
\end{equation*}
where $D_{t,111}^{s,g}=1$. Since the infant death counts are always non-zero, we obtain $D_{t,1}^{s,g}>0$ for each year $t$. CDF enjoys an additional benefit of monotonicity \citep[see also][]{MS13}.

Because of the unit interval constraints of the CDFs, we implement a logistic transformation, given as
\begin{equation*}
X_{t,v}^{s,g} = \text{logit}(D_{t,v}^{s,g}) = \ln \left(\frac{D_{t,v}^{s,g}}{1 - D_{t,v}^{s,g}}\right), \qquad v=1, 2,\dots,110,
\end{equation*}
where $\ln(\cdot)$ denotes the natural logarithm. Since $D_{t,111}^{s,g}=1$, the last column is removed to avoid the undefinedness of the logistic transformation. 

To transform $X_{t,v}^{s,g}$ back to the original scale, we first perform an inverse logit transformation, obtaining
\begin{equation*}
D_{t,v}^{s,g} = \frac{\exp^{X_{t,v}^{s,g}}}{1+\exp^{X_{t,v}^{s,g}}},
\end{equation*}
where we add back the last column of ones. We then take the first-order differencing to obtain
\begin{equation*}
d_{t,x}^{s,g} = \Delta^{x}_{\omega=1}D_{t,\omega}^{s,g} \times 10^5,
\end{equation*}
where $\Delta$ represents the first-order differencing, where $D_{t,1}^{s,g}=d_{t,1}^{s,g}$.

Both transformations provide real-valued data objects, namely $z_{t,x}^{s,g}$ in the clr transformation or $X_{t,v}^{s,g}$ in the CDF transformation. Steps 3) to 4) in the clr transformation present one way to model each series independently of $s$ and $g$. 

In Section~\ref{sec:4}, we present some multi-population functional time-series modeling techniques to take into account relations within and between prefecture $s$ and gender $g$. Within the functional time-series framework, we treat age as the continuum, denoting it by $u\in \mathcal{I}\subset \mathbb{R}$, where the function support $\mathcal{I}\in [0,110+]$ is a subset of real-valued space $\mathbb{R}$.

\section{Multiple functional time-series forecasting models}\label{sec:4}

Because of the dense and equally spaced data design, a functional data-analytic approach provides a convenient notation and embeds a high-dimensional real-valued variable  into a continuous function. Disaggregated by state and gender over time, the age-specific mortality $\X_{t}^{s,g}(u)$ is an example of a high-dimensional functional time series (HDFTS) \citep[see, e.g.,][]{GSY19, TNH23}. In Sections~\ref{sec:4.2} to~\ref{sec:4.5}, we revisit some forecasting method that handle age- and prefecture-specific mortality rates. In Section~\ref{sec:4.4}, we introduce a combination of the two-way functional analysis of variance and functional factor model, which can handle age-, gender- and prefecture-specific mortality rates.

\subsection{Multivariate functional time-series (MFTS) forecasting method}\label{sec:4.2}

For a given gender $g$, we observe $\bm{\X}_t^g(u) = \{\X_t^{1,g}(u),\X_t^{2,g}(u),\dots,\X_t^{S,g}(u)\}$ at $S$ prefectures, where each element is of dimension $T\times M$ and $M$ denotes the number of ages. We consider some multi-population functional time-series forecasting methods \citep[see][for a comprehensive review]{HM24}. Since our multiple functional time series have the same function support, we have data where each observation~$t$ consists of $S$ functions $\{\X_t^{1,g}(u),\X_t^{2,g}(u),\dots,\X_t^{S,g}(u)\}\in~\mathbb{R}^S$, and $S=47$ in our data set.

These multivariate functional time series are stacked in a vector. Let $\bm{\varrho}^g(u)$ be the mean. For any two series $s$ and $j$ with the same function support $u, w\in \mathcal{I}$, the cross-covariance function can be defined with elements
\begin{align*}
C_{sj}(u,w) :=& \ \text{Cov}[\bm{\X}^{s}(u),\bm{\X}^j(w)] \\
		=& \ \text{E}[\bm{\X}^{s}(u) - \bm{\varrho}^{s}(u)][\bm{\X}^j(w)-\bm{\varrho}^j(w)],
\end{align*}
where $\bm{\varrho}^{s}(u)$ and $\bm{\varrho}^j(w)$ denote the mean functions for the $s$\textsuperscript{th} and $j$\textsuperscript{th} series, respectively. The cross-covariance function is of dimension $(SM \times SM)$.

By the separability of Hilbert space, the Karhunen-Lo\`{e}ve expansion of a stochastic process $\X^{s}_t(u)$ can be expressed as
\begin{equation*}
\X_t^{s}(u) \approx \widehat{\varrho}^{s}(u) + \sum^K_{k=1}\widehat{\beta}_{t,k}^{s}\widehat{\phi}_k^{s}(u),
\end{equation*}
where $\widehat{\varrho}^s(u) = \frac{1}{T}\sum^T_{t=1}\X_t^s(u)$, $K$ represents the number of retained functional principal components. Expressing it in a matrix format, 
\begin{equation*}
\bm{\X}_t(u) = \bm{\widehat{\varrho}}(u) + \bm{\widehat{\Phi}}(u)\bm{\widehat{\beta}}_t^{\top},
\end{equation*}
where $\bm{\X}_t(u) = [\X_t^{1}(u),\X_t^2(u),\dots,\X_t^{S}(u)]^{\top}$ and $\bm{\X}(u) = [\bm{\X}_1(u), \bm{\X}_{2}(u),\dots,\bm{\X}_T(u)]^{\top}$ denote stacked historical functions, $\bm{\widehat{\beta}}_t = (\widehat{\beta}_{t,1}^{1},\dots,\widehat{\beta}_{t,K}^{1}, \widehat{\beta}_{t,1}^{2},\dots,\widehat{\beta}_{t,K}^{2},\dots, \widehat{\beta}_{t,1}^{S},\dots, \widehat{\beta}_{t,K}^{S})$ is a vector of the basis expansion coefficients, $\bm{\widehat{\varrho}}(u) = [\widehat{\varrho}^1(u),\widehat{\varrho}^2(u),\dots,\widehat{\varrho}^{S}(u)]^{\top}$ and 
\begin{equation*}
\bm{\widehat{\Phi}}(u) = \begin{pmatrix}
\widehat{\phi}_1^1(u) & \cdots & \widehat{\phi}_K^1(u) & \cdots  \\
 & \cdots & & \widehat{\phi}_1^2(u) & \cdots & \widehat{\phi}_K^2(u) & \cdots \\
 & & & \ddots & & \\
& & &   & & \cdots & \widehat{\phi}_1^{S}(u) & \cdots & \widehat{\phi}_K^{S}(u)  
\end{pmatrix}_{S\times (K\times S)},
\end{equation*}
where $\cdots$ and $\ddots$ show the interaction term between the basis functions of two series.

Conditioning on the past functions and estimated functional principal components $\bm{\widehat{\Phi}}(u)$, the $h$-step-ahead point forecast of $\bm{\X}_{T+h}(u)$ is given as
\begin{equation*}
\widehat{\bm{\X}}_{T+h|T}(u) = \widehat{\bm{\varrho}}(u) + \bm{\widehat{\Phi}}(u)\widehat{\bm{\beta}}_{T+h|T}^{\top},
\end{equation*}
where $\widehat{\bm{\beta}}_{T+h|T}^{\top}$ denotes the univariate time-series forecasts of the functional principal component scores corresponding to the $S$ series. In our implementation, each series is first standardized to have equal scale before being combined into a long vector \citep[see also][]{CCY14}.

\subsection{Multilevel functional time-series (MLFTS) forecasting method}\label{sec:4.3}

The multilevel functional time series model bears a strong resemblance with the two-way functional analysis of variance, studied by many researchers including \cite{MVB+03} and \cite{CF10}. The essential idea is to decompose the HDFTS to a common pattern, a population-specific pattern and residual term. The common and population-specific patterns are obtained via projecting them onto the eigenvectors of covariance functions of the aggregate and population-specific centered stochastic processes, respectively. 

For each gender $g$, the curve at observation~$t$ can be expressed as 
\begin{equation}
\X_t^{s}(u) = \mu^s(u) +  R_t^{c}(u)+U_t^s(u), \label{eq:1}
\end{equation}
where $\widehat{\mu}^s(u)=\frac{1}{T}\sum^{T}_{t=1}\X_t^s(u)$ represents the sample mean of the $s$\textsuperscript{th} series. As a proxy of common pattern without any prior information, one can take the simple average of all series, that is $\widehat{R}_t^{c}(u) = \frac{1}{S}\sum^{S}_{s=1}\X_t^s(u)$. The remaining series-specific pattern is absorbed into the $U_t^s(u)$. 

The sample version of common pattern and population-specific pattern are given by
\begin{alignat}{2}
\widehat{R}^{c}_t(u) &= \sum_{\iota=1}^{T}\widehat{\zeta}_{t,\iota}^{c}\widehat{\psi}_{\iota}^{c}(u)  &&\approx \sum_{\iota=1}^{I}\widehat{\zeta}_{t,\iota}^{c}\widehat{\psi}_{\iota}^{c}(u) \label{eq:4} \\
\widehat{U}_t^s(u) &= \sum_{\xi=1}^{T}\widehat{\tau}_{t,\xi}^{s}\widehat{\vartheta}_{\xi}^s(u) &&\approx \sum^J_{\xi=1}\widehat{\tau}_{t,\xi}^s\widehat{\vartheta}_{\xi}^s(u), \label{eq:3}
\end{alignat}
where $\widehat{\bm{\zeta}}_{\iota}^{c} = (\widehat{\zeta}_{1,\iota}^{c}, \widehat{\zeta}_{2,\iota}^{c}, \dots,\widehat{\zeta}_{T,\iota}^{c})$ for $\iota=1, 2,\dots,I$ are the estimated principal component scores of~$\widehat{\bm{R}}^{c}(u)$, $\widehat{\bm{\tau}}_{\xi}^{s} = (\widehat{\tau}_{1,\xi}^{s}, \widehat{\tau}_{2,\xi}^{s}, \dots,\widehat{\tau}_{T,\xi}^{s})$ for $\xi=1,2,\dots,J$ are the estimated principal component scores of $\widehat{\bm{U}}^s(u)$, $\widehat{\bm{\psi}}^{c}(u) = [\widehat{\psi}_{1}^{c}(u), \widehat{\psi}_{2}^{c}(u), \dots, \widehat{\psi}_{I}^{c}(u)]$ are the estimated functional principal components of $\widehat{\bm{R}}^{c}(u)$, and $\widehat{\bm{\vartheta}}^s(u) = [\widehat{\vartheta}_{1}^s(u), \widehat{\vartheta}_{2}^s(u),\dots, \widehat{\vartheta}_{J}^s(u)]$ are the estimated functional principal components of~$\widehat{\bm{U}}^s(u)$.

Substituting~\eqref{eq:4} and~\eqref{eq:3} into~\eqref{eq:1}, we obtain
\begin{equation*}
\X_t^{s}(u) = \widehat{\mu}^s(u) + \sum_{\iota=1}^{I}\widehat{\zeta}_{t,\iota}^{c}\widehat{\psi}_{\iota}^{c}(u) + \sum^J_{\xi=1}\widehat{\tau}_{t,\xi}^s\widehat{\vartheta}_{\xi}^s(u) + e_t^s(u),
\end{equation*}
where $e_t^s(u)$ denotes measurement error with a finite variance.

To select the optimal number of components, we consider an eigenvalue ratio criterion of \cite{LRS20} to determine $I$ and $J$. In essence, the optimal number of components is determined by having the smallest ratio between two successive eigenvalues in~\eqref{eq:EVR}. As an alternative recommended in \cite{HBY13}, we set $I=J=6$.

Conditioning on observed data $\bm{\X}^s(u) = [\X_1^s(u), \X_2^s(u), \dots,\X_T^s(u)]$, estimated series-specific mean $\widehat{\mu}^s(u)$, estimated principal components $\widehat{\bm{\psi}}^{c}(u)$ and $\widehat{\bm{\vartheta}}^s(u)$, the $h$-step-ahead point forecasts can be obtained by
\begin{align*}
\widehat{\X}^s_{T+h|T}(u) &= \text{E}\left[\X_{T+h}^s(u)|\bm{\X}^s(u), \widehat{\mu}^s(u), \widehat{\bm{\psi}}^{c}(u), \widehat{\bm{\vartheta}}^s(u)\right] \\
&= \widehat{\mu}^s(u) + \sum^{I}_{\iota=1}\widehat{\zeta}_{T+h|T,\iota}^{c}\widehat{\psi}_{\iota}^{c}(u) + \sum^J_{\xi=1}\widehat{\tau}_{T+h|T,\xi}^s\widehat{\vartheta}_{\xi}^s(u),
\end{align*}
where $\widehat{\zeta}_{T+h|T,\iota}^{c}$ and $\widehat{\tau}_{T+h|T,\xi}^s$ are the forecasted principal component scores, obtained from a univariate time-series forecasting method, such as autoregressive integrated moving average (ARIMA) or~ETS.

\subsection{Two-stage functional principal component analyses}\label{sec:4.5}

\cite{GSY19} put forward a two-stage approach combining principal component analysis and a multivariate factor model for the resulting panels of principal component scores. The method can be summarized into the following steps:
\begin{enumerate}[1)]
\item Implement a dynamic functional principal component on each set of functional time series, resulting in $S$ sets of principal component scores of low dimension $p_0$.
\item The first principal component scores from each of $S$ series are combined into an $S\times 1$ vector. Then, a factor model is fitted to the vector of scores to further reduce the dimension into an $r\times 1$ vector, where $r<<S$. The same procedure is implemented for the second, third, and so on, until the $p_0$\textsuperscript{th} scores. The vector of $S$ series is reduced to an $r\times p_0$ matrix.
\item A univariate time series model is then fitted to each factor, and forecasts are produced. The forecast factors can then be used to construct forecast HDFTS.
\end{enumerate}

\subsection{Two-way functional analysis of variance}\label{sec:4.4}

Through a two-way FANOVA, the clr- or CDF-transformed data $\X_{t}^{s,g}(u)$ can be decomposed into
\begin{equation*}
\X_{t}^{s,g}(u) = \mu(u)+\alpha^s(u)+\delta^g(u)+\gamma^{s,g}(u)+\Y_t^{s,g}(u),
\end{equation*}
where $\mu(u)$ represents the functional grand effects, $\alpha^s(u)$ denotes the functional row (prefecture) effect, $\delta^g(u)$ signifies the functional column (gender) effect, $\gamma^{s,g}(u)$ denotes the interaction term, and $\Y_{t}^{s,g}(u)$ represents the residual component. Since our goal is forecasting rather than inference, we group the interaction term and the residual component, and denote 
\begin{equation*}
\Z_{t}^{s,g}(u) = \gamma^{s,g}(u)+\Y_{t}^{s,g}(u).
\end{equation*}

The functional grand, row, and column effects can be estimated by their sample means \citep[see, e.g.,][Chapter 13]{RS06}. They are given by
\begin{align}
\widehat{\mu}(u) &= \frac{1}{S\times 2\times T}\sum^{S}_{s=1}\sum^2_{g=1}\sum^T_{t=1}\X_t^{s,g}(u) \tag{grand mean}\\
\widehat{\alpha}^s(u) &= \frac{1}{2\times T}\sum^{2}_{g=1}\sum^T_{t=1}\X_t^{s,g}(u) - \widehat{\mu}(u) \tag{prefecture mean}\\
\widehat{\delta}^g(u) &= \frac{1}{S\times T}\sum^{S}_{s=1}\sum^T_{t=1}\X_t^{s,g}(u) - \widehat{\mu}(u) \tag{gender mean}
\end{align}
To ensure identifiability, some constraints are in place:
\begin{equation*}
\sum^{S}_{s=1}\alpha^s(u) = \sum^2_{g=1}\delta^g(u) = 0,\qquad \sum^{S}_{s=1}\Z_{t}^{s,g}(u) = \sum^2_{g=1}\Z_t^{s,g}(u) = 0, \quad \forall t.
\end{equation*}

The estimates $\widehat{\mu}(u)$, $\widehat{\alpha}^s(u)$, $\widehat{\delta}^g(u)$ capture the grand mean, prefecture mean, and gender mean, they are highly interpretable. The residual component $\Z_{t}^{s,g}(u)$ is time-varying and itself is an HDFTS. To model $\Z_t^{s,g}(u)$, we resort to a functional factor model.

It is important to acknowledge that several variants of functional factor models exist in the literature. For example, \cite{HNT23} and \cite{TNH23} present a factor model with functional factor loadings and scalar-valued factors. \cite{GQW25} propose a factor model with scalar-valued factor loadings and functional factors. By unifying both factor models, \cite{LLS+26} introduce a factor model with functional factor loadings and functional factors. 

\paragraph {Functional factor model:} Since females and males differ biologically, we consider each gender separately. For each gender $g$, we consider the following functional factor model:
\begin{equation}
\Z_{t}^{s,g}(u) = \sum^{\mathcal{K}}_{\mathrm{k}=1}\int_{\mathcal{I}_{\mathrm{k}}}B_{\mathrm{k}}^{s,g}(u,v)F_{t,\mathrm{k}}^{g}(v)dv + \varepsilon_t^{s,g}(u),\qquad u\in \mathcal{I}_s,\label{eq:6}
\end{equation}
where $B_{\mathrm{k}}^{s,g}(u,v)$ represents the $\mathrm{k}$\textsuperscript{th} two-dimensional functional loading, $F_{t,k}^g(v)$ represents time-varying functional factors, $\mathcal{K}$ is the number of factors, $\varepsilon_{t}^{s,g}(u)$ is the model error term, $\mathcal{I}_s$ denotes the function support range for each prefecture $s=1, 2,\dots,S$, and $\mathcal{I}_{\mathrm{k}}$ represents a compact set which may be different from $\mathcal{I}_s$. By imposing a low-dimensional functional factor condition on the latent factor $F_{t,\mathrm{k}}^g(v)$, a series approximation is
\begin{equation}
F_{t,\mathrm{k}}^g(v) = \Phi_{\mathrm{k}}^g(v)^{\top}\bm{G}_t^g + \eta_{t,\mathrm{k}}^g(v),\qquad v\in \mathcal{I}_k,\label{eq:5}
\end{equation}
where $\Phi_{\mathrm{k}}^g(v)$ is a $q$-dimensional vector of basis functions, $\bm{G}_t^g$ is a $q$-dimensional vector of random variables, and $\eta_{t,\mathrm{k}}^g(v)$ is the series approximation error.

By plugging~\eqref{eq:5} into~\eqref{eq:6}, we obtain
\begin{align*}
\Z_{t}^{s,g}(u) &= \sum^{\mathcal{K}}_{\mathrm{k}=1}\int_{\mathcal{I}_{\mathrm{k}}}B_{\mathrm{k}}^{s,g}(u,v)\left[\Phi_{\mathrm{k}}^g(v)^{\top}\bm{G}_t^g + \eta_{t,\mathrm{k}}^g(v)\right]dv + \varepsilon_t^{s,g}(u) \\
&= \sum^{\mathcal{K}}_{\mathrm{k}=1}\int_{\mathcal{I}_{\mathrm{k}}}B_{\mathrm{k}}^{s,g}(u,v)\Phi_{\mathrm{k}}^g(v)^{\top}dv\bm{G}_t^g + \sum^{\mathcal{K}}_{\mathrm{k}=1}\int_{\mathcal{I}_{\mathrm{k}}}B_{\mathrm{k}}^{s,g}(u,v)\eta_{t,\mathrm{k}}^g(v)dv + \varepsilon_t^{s,g}(u) \\
&=\Lambda^{s,g}(u)^{\top}\bm{G}_t^g + \epsilon_{t}^{s,g}(u),
\end{align*}
where $\Lambda^{s,g}(u) = \sum^{\mathcal{K}}_{\mathrm{k}=1}\int_{\mathcal{I}_{\mathrm{k}}}B_{\mathrm{k}}^{s,g}(u,v)\Phi_{\mathrm{k}}^g(v)dv$ denotes an aggregation of $\mathcal{K}$ factors, and the error is denoted as $\epsilon_{t}^{s,g}(u)=\sum^{\mathcal{K}}_{\mathrm{k}=1}\int_{\mathcal{I}_{\mathrm{k}}}B_{\mathrm{k}}^{s,g}(u,v)\eta_{t, \mathrm{k}}^g(v)dv + \varepsilon_t^{s,g}(u)$.

To estimate $\bm{G}_t^g$, we first estimate the covariance of $\Z_{t}^{s,g}(u)$ by
\begin{equation*}
\bm{\nabla}^g = (\nabla^g_{tt^{'}})_{T\times T} \qquad \text{with}\qquad \nabla_{tt^{'}}^g = \frac{1}{S}\sum_{s=1}^{S}\int_{u\in \mathcal{I}_s}\Z_{t}^{s,g}(u)\Z_{t^{'}}^{s,g}(u)du.
\end{equation*}

Applying eigen-decomposition to the $T\times T$ matrix $\bm{\nabla}^g$, we obtain a set of normalized $\widetilde{\bm{G}}^g = (\widetilde{\bm{G}}_1^g,\widetilde{\bm{G}}_2^g,\dots,\widetilde{\bm{G}}_T^g)^{\top}$ as a $T\times q$ matrix with columns being the eigenvectors (multiplied by $\sqrt{T}$) corresponding to the $q$ largest eigenvalues of $\bm{\nabla}^g$. Via ordinary least squares, the factor loadings are then estimated as
\begin{equation*}
\Lambda^{s,g}(u) = \frac{1}{T}\sum^{T}_{t=1}\Z_t^{s,g}(u)\widetilde{\bm{G}}_t^g,
\end{equation*}
with a normalization constraint $\frac{1}{T}\sum^T_{t=1}\widetilde{\bm{G}}_t^g(\widetilde{\bm{G}}_t^{g})^{\top}=\bm{I}_q^g$, where $q$ is the dimensionality of the low-rank representation in~\eqref{eq:5}.

To estimate the order $q$ for gender~$g$, \cite{LLS+26} introduced an information criterion. Let $\delta_{\ell}(\bm{\nabla}^g/T)$ be the $\ell$\textsuperscript{th} largest eigenvalue of $\bm{\nabla}^g/T$ and define
\begin{equation*}
\widehat{q}^g = \argmin_{1\leq \ell \leq T}\left[\delta_{\ell}(\bm{\nabla}^g/T)+\ell \cdot \phi_{ST}\right]-1,
\end{equation*}
where $\phi_{ST}$ is the penalty term, which we set $\phi_{ST}=[\max(S, T)]^{-\frac{1}{2}}$.

Conditonal on the estimated factor loadings $\widetilde{\Lambda}^{s,g}(u)$, the time-varying dynamics are captured by the $q$-dimensional factors $\widetilde{\bm{G}}_t^g$. By applying a univariate time-series forecasting method, such as ETS, we obtain its $h$-step-ahead forecast of $\widehat{\widetilde{\bm{G}}}_{T+h|T}^{g}$. The order of ETS is chosen by an automatic algorithm, based on the corrected Akaike information criterion of \cite{HT93}.

By multiplying the forecasted factors with the estimated factor loadings, the $h$-step-ahead forecast of $\Z_{T+h}^{s,g}(u)$ is given by
\begin{equation*}
\widehat{\Z}_{T+h|T}^{s,g}(u) = \Lambda^{s, g}(u)^{\top}\widehat{\widetilde{\bm{G}}}_{T+h|T}^{g},
\end{equation*}
where $\widehat{\widetilde{\bm{G}}}_{T+h|T}^{g}$ denotes the $h$-step-ahead time-series forecast of $\widetilde{\bm{G}}_{T+h}^{g}$.

\section{Gap modeling and forecasting}\label{sec:5}

\subsection{Gender gap}\label{sec:5.1}

\begin{figure}[!htb]
\centering
\includegraphics[width=5.91cm]{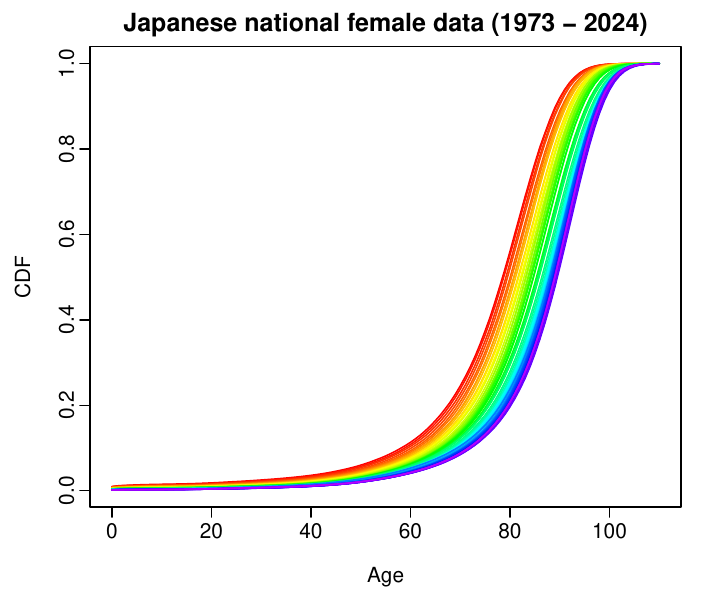}
\includegraphics[width=5.91cm]{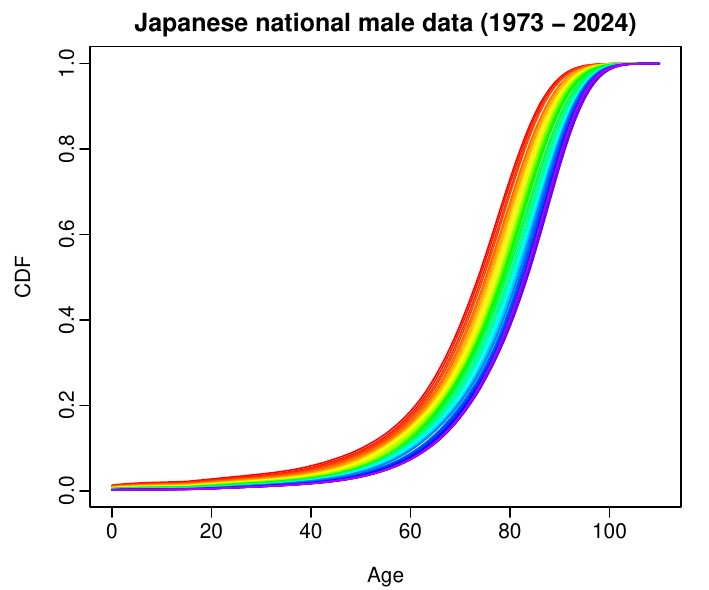}
\includegraphics[width=5.91cm]{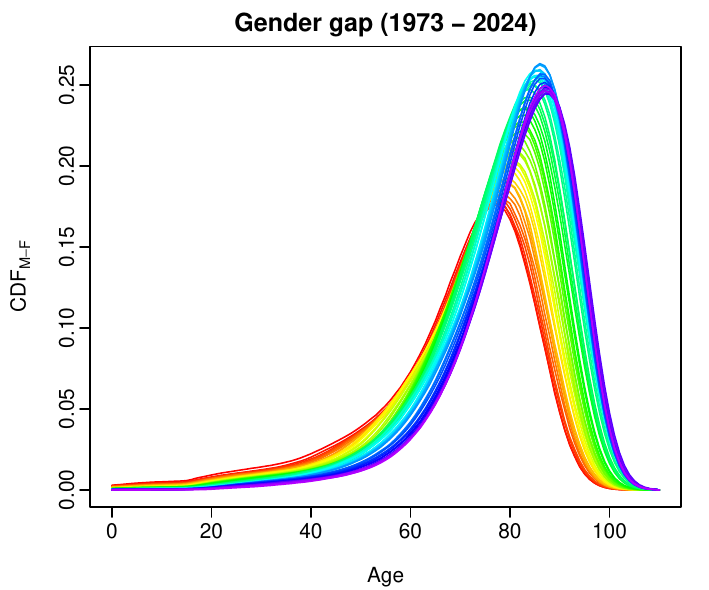}
\caption{\small Cumulative distribution functions of Japanese female and male relative life-table death counts from 1973 to 2024, and their gender gap.}\label{fig:2}
\end{figure}

Gender differences in mortality are a long-standing and widely documented feature of population health, reflecting biological, behavioural and social factors \citep[see, e.g.,][]{LAO+13}. The extensively studied pattern is the systematic survival advantage of females over males, observed globally \citep[see, e.g.,][]{LG14} and Europe \citep[see, e.g.,][]{GH07, ZJO+21}. Like many developed countries, males in Japan experience higher cumulative relative life-table death counts than females. Among centenarians, the ratio of females to males is about 7:1. Based on the national or subnational cumulative relative life-table death counts, we compute the gender gap as
\begin{equation*}
G_{t,x}^{\text{N}, \text{M}-\text{F}} = D_{t,x}^{\text{N}, \text{M}} - D_{t,x}^{\text{N}, \text{F}}, \qquad G_{t,x}^{s, \text{M}-\text{F}} = D_{t,x}^{s, \text{M}} - D_{t,x}^{s, \text{F}}, \qquad s=1,2,\dots,S,
\end{equation*}
an example could be the national gender gap shown in Figure~\ref{fig:2}.

Through~\eqref{eq:PCA} and~\eqref{eq:2}, we separately model and forecast the time series of age-specific gender gap and the time series of age-specific female life-table death counts. By adding the $h$-step-ahead forecast of the female data and gender gap, we obtain the $h$-step-ahead forecast of the male data.

\subsection{Region gap}\label{sec:5.2}

Region heterogeneity reflects differences between regional and national data. Some prefectures, such as Okinawa, are known to have more centenarians than others \citep{WWH+08, PH24}. Such heterogeneity presents in-depth geographical understanding of mortality differentials and offers valuable information on how regional distinctions, such as health policies, education attainment, socioeconomic status, can influence mortality between regions \citep[see, e.g.,][]{BD20, BPB23}. Let the difference between the subnational and national data for a given gender $g$ be represented by
\begin{equation*}
U_{t,x}^{s-\text{N},g} =D_{t,x}^{s,g} - D_{t,x}^{\text{N},g}, \qquad g\in \{\text{F}, \text{M}\}.
\end{equation*}

Through~\eqref{eq:PCA} and~\eqref{eq:2}, we separately model and forecast the time series of age-specific region gap and the time series of age-specific national life-table death counts. By adding the $h$-step-ahead forecast of the national data and region gap, we obtain the $h$-step-ahead forecast of the subnational data for a particular gender.

\subsection{Double gap}\label{sec:5.3}

By combining the region and gender gaps, we consider the double gap idea. Via principal component regression in~\eqref{eq:PCA} and~\eqref{eq:2}, we first model and forecast the national female life-table death counts and the region gap; in doing so, we obtain the subnational female forecasts. Then, we model and forecast gender gap, from which we obtain subnational male forecasts. The gap modeling is not limited to the CDF transformation, it also works for the clr transformation with one subtle difference: the zero replacement described in~\eqref{eq:zero_replacement}. 
                        
\section{Construction of pointwise prediction intervals} \label{sec:6}

\begin{figure}[!htb]
\centering
\begin{tikzpicture}[font=\small, period/.style={draw, minimum height=0.8cm, align=center}]
\node[period, fill=gray!35, minimum width=3.0cm] at (1.5,0) {Training};
\node[period, fill=gray!20, minimum width=5.5cm] at (5.75,0) {Validation};
\node[period, fill=gray!10, minimum width=5.0cm] at (11,0) {Test};
\node[below] at (0,-0.55) {1973};
\node[below] at (3,-0.55) {1983};
\node[below] at (8.5,-0.55) {2004};
\node[below] at (13.5,-0.55) {2024};
\end{tikzpicture}
\caption{\small Partition of the sample into training (1973--1983), validation (1984--2004) and test (2005--2024) periods.}\label{fig:sample_partition}
\end{figure}
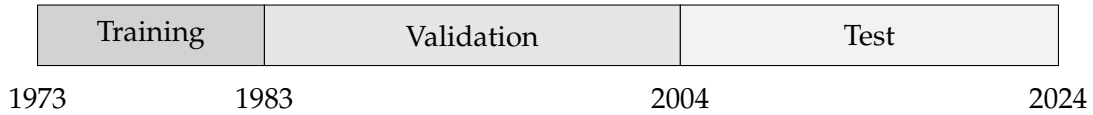

We partition the 52-year data sample (1973--2024) into training, validation and test sets with proportions of about 20\%, 40\% and 40\%, respectively. Using the initial training period from 1973 to 1983, we adopt an expanding-window forecasting scheme to generate $h$-step-ahead forecasts for the validation period 1984--2004, for $h=1, 2,\dots, 20$. In Figure~\ref{fig:sample_partition}, we display the years associated with the training, validation and test sets.

We implement an expanding-window forecast scheme to evaluate and compare point forecast accuracy among the multi-population and gap modeling approaches. The expanding-window scheme grows sample size of the training sample iteratively. The number of available curves in the validation or test set varies with the forecast horizon $h$: for instance, when $h=1$, there are 21 years to compute residual function, which is the difference between the observed curves in the validation set and their forecasts; when $h=20$, only two years are available. From these residual functions, we compute pointwise summary measure $\chi^{s,g}(u_i)$ for each prefecture $s$ and gender $g$, such as the pointwise standard deviation. 

For a given horizon $h$, let the residual functions be denoted as $\upsilon_{\varpi}^{s,g}(u_i) = d_{\varpi}^{s,g}(u_i) - \widehat{d}_{\varpi}^{s,g}(u_i)$ for $\varpi=1, 2,  \dots, T_{\nu}$, and $T_{\nu}$ denotes the number of years in the validation set. Our objective is to determine a tuning parameter $\rho_{\alpha}$ such that $100(1-\alpha)\%$ of the residuals satisfy
\begin{equation*}
-\rho_{\alpha}\chi^{s,g}(u_i) \leq \upsilon_{\varpi}^{s,g}(u_i) \leq \rho_{\alpha}\chi^{s,g}(u_i), \qquad i=1,2,\dots,111.
\end{equation*}
When $T_{\nu}$ is relatively large, one could achieve
\begin{small}
\begin{equation*}
\mathbb{P}\left[-\rho_{\alpha}\chi^{s,g}(u_i) \leq d_{T+h}^{s,g}(u_i) - \widehat{d}_{T+h|T}^{s,g}(u_i) \leq \rho_{\alpha}\chi^{s,g}(u_i)\right]\approx \frac{1}{T_{\nu}}\sum^{T_{\nu}}_{\varpi=1}\mathds{1}\left\{-\rho_{\alpha}\chi^{s,g}(u_i)\leq \upsilon_{\varpi}^{s,g}(u_i) \leq \rho_{\alpha}\chi^{s,g}(u_i) \right\},
\end{equation*}
\end{small}
where $d_{T+h}^{s,g}(u_i)$ denotes the age-specific life-table death count at time $T+h$ in prefecture~$s$ and gender~$g$, and $\widehat{d}_{T+h|T}^{s,g}(u_i)$ denotes the corresponding time-series forecasts at the original scale.

\section{Results}\label{sec:7}

\subsection{Expanding-window scheme}\label{sec:7.1}

\begin{figure}[!htb]
\begin{center}
\begin{tikzpicture}
\draw[->] (0,0) -- (10,0) node[right] {Time};
\draw[fill=blue!20] (0,-0.5) rectangle (3,0.5) node[midway] {Train};
\draw[fill=red!20] (3,-0.5) rectangle (3.5,0.5) node[midway] {F};
\draw[fill=blue!20] (0,-1.5) rectangle (5,-0.5) node[midway] {Train};
\draw[fill=red!20] (5,-1.5) rectangle (5.5,-0.5) node[midway] {F};
\draw[fill=blue!20] (0,-2.5) rectangle (7,-1.5) node[midway] {Train};
\draw[fill=red!20] (7,-2.5) rectangle (7.5,-1.5) node[midway] {F};
\draw[fill=blue!20] (0,-3.5) rectangle (9,-2.5) node[midway] {Train};
\draw[fill=red!20] (9,-3.5) rectangle (9.5,-2.5) node[midway] {F};
\node[left] at (0,0) {1973:2004};
\node[left] at (0,-1) {1973:2005};
\node[left] at (0,-2) {\hspace{-0.8in}{$\vdots$}};
\node[left] at (0,-3) {1973:2023};
\draw[fill=blue!20] (6.5,1) rectangle (7,1.5);
\node[right] at (7,1.25) {Training Window};
\draw[fill=red!20] (6.5,0.5) rectangle (7,1);
\node[right] at (7,0.75) {Forecast (F)};
\end{tikzpicture}
\end{center}
\caption{\small A diagram of the expanding-window forecast scheme. The data begin in 1973 and end in 2024. We demonstrate one-step-ahead prediction and also consider other horizons up to 20 to evaluate point and interval forecast accuracies.}\label{fig:expanding}
\end{figure}
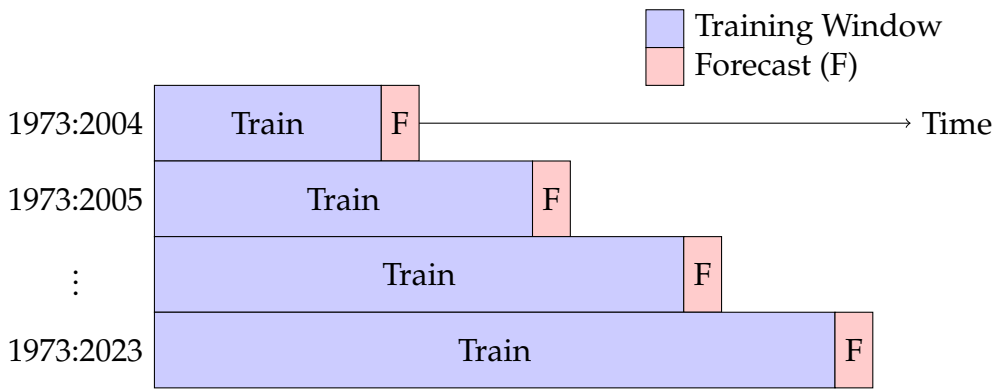

With an expanding-window scheme, let the initial training sample be the subnational age-specific life-table death counts from 1973 to 2004, consisting of the first 32 years of observations. For a given horizon $h$, we compute the $h$-step-ahead point forecast. Then, we increase the training sample size by one year, and re-compute the $h$-step-ahead point forecast. This recursive procedure yields a total of 20 one-step-ahead forecasts, 19 two-step-ahead forecasts, and one 20-step-ahead forecast. Figure~\ref{fig:expanding} illustrates this expanding-window scheme for forecast horizon $h=1$, through horizons up to $h=20$ are evaluated.

\subsection{Point forecast evaluation metric}\label{sec:7.2}

Since the life-table death counts resemble a PDF, we consider some density estimation measures to evaluate the point forecast accuracy. These metrics include the discrete version of the KLD \citep{KL51}. KLD is used to measure the loss of information between the estimated life-table death count $\widehat{d}_{T+h}^{s,g}(u)$ and holdout data $d_{T+h}^{s,g}(u)$. The discrete version of the symmetric KLD is defined as
\begin{align}
\text{KLD}^{s,g}_h =\ & D_{\text{KL}}\left[d_{T_{\nu}+\omicron}^{s,g}(u_i)||\widehat{d}_{T_{\nu}+\omicron}^{s,g}(u_i)\right] + D_{\text{KL}}\left[\widehat{d}_{T_{\nu}+\omicron}^{s,g}(u_i)||d_{T_{\nu}+\omicron}^{s,g}(u_i)\right] \notag\\
= \ & \frac{1}{111\times (21-h)}\sum^{20}_{\omicron=h}\sum^{111}_{i=1} d_{T_{\nu}+\omicron}^{s,g}(u_i)\cdot \left[\ln d_{T_{\nu}+\omicron}^{s,g}(u_i) - \ln \widehat{d}_{T_{\nu}+\omicron}^{s,g}(u_i)\right] + \notag\\
\ & \frac{1}{111\times (21-h)}\sum^{20}_{\omicron=h}\sum^{111}_{i=1} \widehat{d}_{T_{\nu}+\omicron}^{s,g}(u_i)\cdot \left[\ln \widehat{d}_{T_{\nu}+\omicron}^{s,g}(u_i) - \ln d_{T_{\nu}+\omicron}^{s,g}(u_i)\right],\label{eq:KLD}
\end{align}
where $u_i$ denotes a discrete age, and $T_{\nu}$ represents the base year from which forecasts are computed. By averaging across all horizons, the averaged KLD, denoted by $\overline{\text{KLD}}^{s,g}$, can be computed as
\begin{equation*}
\overline{\text{KLD}}^{s,g} = \frac{1}{20}\sum^{20}_{h=1}\text{KLD}^{s,g}_h.
\end{equation*}

\subsection{Point forecast comparison}\label{sec:7.3}

In Table~\ref{tab:1}, using the ETS forecasting method, we report the KLD values for selected forecast horizons, together with overall summary statistics for the multi-population and gap modeling methods. For the female data, HDFPCA achieves the smallest average KLD. For the male data, FANOVA+FFM yields the smallest average KLD. Fixing $K=6$ helps the univariate method but can badly hurt the multivariate ones; EVR should therefore be preferred where the two disagree. The KLD generally increases as the forecast horizon~$h$ increases, reflecting the greater uncertainty associated with longer-horizon forecasts. Between the two transformations, the clr transformation offers better performance, particularly when combined with the ETS forecasting method.
\begin{small}
\tabcolsep 0.037in
\renewcommand{\arraystretch}{1}
\begin{longtable}{@{}llrrrrrrrrrrr@{}}
\caption{\small Averaged over 47 prefectures, we evaluate and compare the point forecast accuracy measured by the KLD under the CDF and clr transformations. For several methods, we compare the results under the EVR and $K=6$. Due to the limited space, we present $h=1, 5, 10, 15, 20$. The univariate time-series forecasting method is ETS. The best method in each summary row is bolded. FANOVA+FFM represents the combination of two-way FANOVA and FFM.}\label{tab:1}\\
\toprule
& & \multicolumn{3}{c}{EVR} & & FANOVA & \multicolumn{3}{c}{$K=6$} & \multicolumn{3}{c}{Gap} \\
\cmidrule(lr){3-5}\cmidrule(lr){8-10}\cmidrule(lr){11-13}
Sex & $h$ & UFTS & MFTS & MLFTS & HDFPCA & FFM &
UFTS & MFTS & MLFTS & Gender & Region & Double \\
\midrule
\endfirsthead
\toprule
& & \multicolumn{3}{c}{EVR} & & FANOVA & \multicolumn{3}{c}{$K=6$} & \multicolumn{3}{c}{Gap} \\
\cmidrule(lr){3-5}\cmidrule(lr){8-10}\cmidrule(lr){11-13}
Sex & $h$ & UFTS & MFTS & MLFTS & HDFPCA & FFM &
UFTS & MFTS & MLFTS & Gender & Region & Double \\
\midrule
\endhead
\midrule
\multicolumn{13}{r}{Continued on next page}\\
\endfoot
\bottomrule
\endlastfoot
\multicolumn{5}{l}{\hspace{-0.1in} \underline{CDF transformation (ETS)}} & \\
F 	& 1 & 0.0053 & 0.0099 & 0.0051 & 0.0053 & 0.0049 & 0.0050 & 0.0059 & 0.0050 & 0.0053 & 0.0043 & 0.0043 \\ 
	& 5 & 0.0111 & 0.0165 & 0.0091 & 0.0075 & 0.0083 & 0.0107 & 0.0112 & 0.0093 & 0.0110 & 0.0071 & 0.0071 \\ 
	& 10 & 0.0284 & 0.0260 & 0.0196 & 0.0178 & 0.0129 & 0.0275 & 0.0227 & 0.0207 & 0.0279 & 0.0124 & 0.0124 \\ 
	& 15 & 0.0766 & 0.0431 & 0.0482 & 0.0183 & 0.0286 & 0.0752 & 0.0480 & 0.0506 & 0.0753 & 0.0316 & 0.0316 \\ 
	& 20 & 0.2466 & 0.1283 & 0.2019 & 0.0237 & 0.0810 & 0.2449 & 0.1631 & 0.1911 & 0.2436 & 0.0840 & 0.0840 \\ 
\cmidrule{2-13}
	& Mean & 0.0614 & 0.0388 & 0.0447 & \textBF{0.0166} & 0.0233 & 0.0605 & 0.0411 & 0.0448 & 0.0611 & 0.0249 & 0.0249 \\ 
	& Median & 0.0314 & 0.0278 & 0.0216 & 0.0155 & 0.0136 & 0.0305 & 0.0249 & 0.0228 & 0.0309 & \textBF{0.0133} & \textBF{0.0133} \\ 
\midrule
M 	& 1 & 0.0046 & 0.0052 & 0.0040 & 0.0043 & 0.0046 & 0.0043 & 0.0040 & 0.0037 & 0.0066 & 0.0035 & 0.0043 \\ 
	& 5 & 0.0077 & 0.0085 & 0.0065 & 0.0080 & 0.0077 & 0.0074 & 0.0071 & 0.0064 & 0.0206 & 0.0059 & 0.0122 \\ 
	& 10 & 0.0134 & 0.0134 & 0.0107 & 0.0160 & 0.0096 & 0.0132 & 0.0121 & 0.0112 & 0.0632 & 0.0109 & 0.0338 \\ 
	& 15 & 0.0189 & 0.0159 & 0.0174 & 0.0418 & 0.0106 & 0.0189 & 0.0165 & 0.0173 & 0.1476 & 0.0158 & 0.0633 \\ 
	& 20 & 0.0697 & 0.0357 & 0.0725 & 0.1159 & 0.0234 & 0.0698 & 0.0464 & 0.0693 & 0.4415 & 0.0978 & 0.1342 \\ 
\cmidrule{2-13}
	& Mean & 0.0176 & 0.0146 & 0.0178 & 0.0322 & \textBF{0.0105} & 0.0175 & 0.0150 & 0.0175 & 0.1155 & 0.0177 & 0.0461 \\ 
	& Median & 0.0140 & 0.0136 & 0.0112 & 0.0174 & \textBF{0.0098} & 0.0138 & 0.0128 & 0.0118 & 0.0697 & 0.0115 & 0.0368 \\ 
\midrule
\multicolumn{5}{l}{\hspace{-0.1in} \underline{clr transformation (ETS)}} & \\
F 	& 1 & 0.1031 & 0.0066 & 0.0086 & 0.0074 & 0.0045 & 0.0122 & 0.0056 & 0.0068 & 0.0052 & 0.0045 & 0.0045 \\ 
	& 5 & 0.1226 & 0.0088 & 0.0235 & 0.0095 & 0.0065 & 0.0318 & 0.0100 & 0.0121 & 0.0103 & 0.0070 & 0.0070 \\ 
	& 10 & 0.1505 & 0.0091 & 0.0222 & 0.0206 & 0.0086 & 0.0993 & 0.0271 & 0.0240 & 0.0232 & 0.0111 & 0.0111 \\ 
	& 15 & 0.1811 & 0.0110 & 0.1020 & 0.0165 & 0.0213 & 0.2548 & 0.0390 & 0.0680 & 0.0572 & 0.0184 & 0.0184 \\ 
	& 20 & 0.2020 & 0.0100 & 0.2567 & 0.0163 & 0.0540 & 0.5956 & 0.1529 & 0.2768 & 0.1720 & 0.0530 & 0.0530 \\ 
\cmidrule{2-13}
	& Mean & 0.1533 & \textBF{0.0093} & 0.0648 & 0.0161 & 0.0163 & 0.1620 & 0.0412 & 0.0581 & 0.0464 & 0.0175 & 0.0175 \\ 
	& Median & 0.1534 & 0.0092 & 0.0299 & 0.0164 & \textBF{0.0089} & 0.1048 & 0.0323 & 0.0263 & 0.0257 & 0.0114 & 0.0114 \\ 
\cmidrule{2-13}
M 	& 1 & 0.0551 & 0.0043 & 0.0065 & 0.0045 & 0.0042 & 0.0066 & 0.0040 & 0.0088 & 0.0058 & 0.0037 & 0.0056 \\ 
	& 5 & 0.0686 & 0.0070 & 0.0109 & 0.0082 & 0.0057 & 0.0155 & 0.0068 & 0.0217 & 0.0098 & 0.0060 & 0.0086 \\ 
	& 10 & 0.0882 & 0.0111 & 0.0209 & 0.0122 & 0.0061 & 0.0407 & 0.0127 & 0.0442 & 0.0173 & 0.0100 & 0.0134 \\ 
	& 15 & 0.1049 & 0.0143 & 0.0679 & 0.0212 & 0.0072 & 0.0807 & 0.0146 & 0.0525 & 0.0330 & 0.0146 & 0.0124 \\ 
	& 20 & 0.1205 & 0.0133 & 0.2170 & 0.1631 & 0.0161 & 0.0418 & 0.0804 & 0.2482 & 0.0972 & 0.1040 & 0.0239 \\ 
\cmidrule{2-13}
	& Mean & 0.0876 & 0.0104 & 0.0533 & 0.0229 & \textBF{0.0072} & 0.0384 & 0.0172 & 0.0745 & 0.0284 & 0.0174 & 0.0128 \\ 
	& Median & 0.0901 & 0.0115 & 0.0242 & 0.0128 & \textBF{0.0063} & 0.0358 & 0.0132 & 0.0438 & 0.0185 & 0.0103 & 0.0128 \\ 
\end{longtable}
\end{small}

\begin{figure}[!htb]
\centering
{\includegraphics[width=8.71cm]{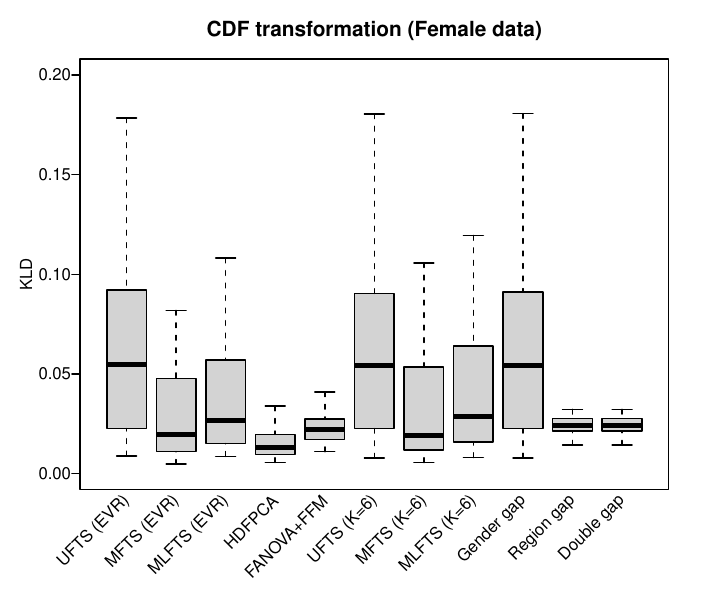}}
\quad
{\includegraphics[width=8.71cm]{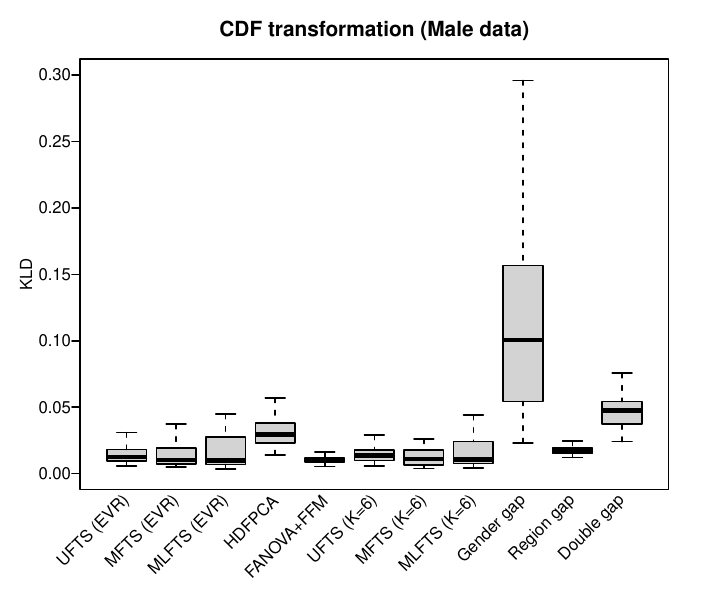}}
\\
{\includegraphics[width=8.71cm]{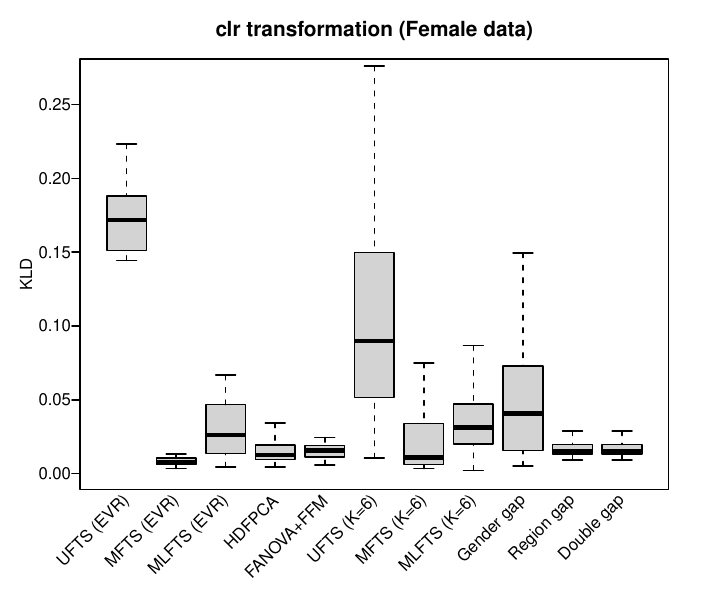}}
\quad
{\includegraphics[width=8.71cm]{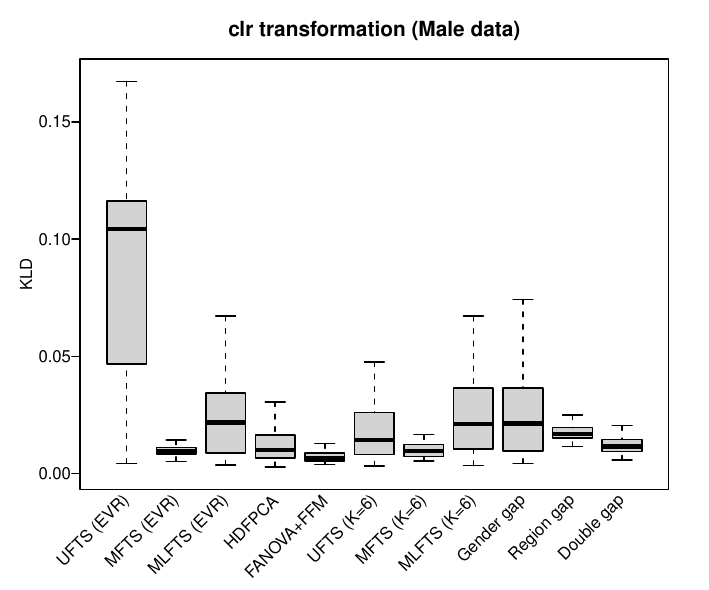}}
\caption{\small Using the $\overline{\text{KLD}}^{s,g}$ as the measure of point forecast accuracy, we compare the performance of multi-population and gap modeling under the ETS forecasting method, considering both the CDF and clr transformations.}\label{fig:3}
\end{figure}

In Figure~\ref{fig:3}, we display the $\overline{\text{KLD}}^{s,g}$ for the female and male data under the CDF and clr transformations. For the female data, the HDFPCA provides the smallest errors under the CDF transformation, while the MFTS (EVR) provides the smallest errors under the clr transformation. For the male data,

\subsection{Interval forecast evaluation metric}\label{sec:7.4}

To evaluate and compare interval forecast accuracy, we consider evaluating empirical coverage probability (ECP), coverage probability difference (CPD), and mean interval score (MIS) of \cite{GR07, GK14}. For each year in the test set, the $h$-step-ahead prediction intervals are calculated at the $(1-\alpha)$ nominal coverage probability; let $\widehat{d}_{T+h,\text{lb}}^{s,g}(u_i)$ and $\widehat{d}_{T+h,\text{ub}}^{s,g}(u_i)$ be the lower and upper bounds, respectively. For a prefecture $s$ and gender $g$, the ECP and CPD are defined as
\begin{align*}
\text{ECP}_{\alpha,h}^{s,g} = \frac{1}{111\times (21-h)}\sum^{20}_{\omicron = h}\sum^{111}_{i=1}&\mathds{1}\left\{\widehat{d}_{T_{\nu}+\omicron, \text{lb}}^{s,g}(u_i)\leq d_{T_{\nu}+\omicron}^{s,g}(u_i)\leq \widehat{d}_{T_{\nu}+\omicron, \text{ub}}^{s,g}(u_i)\right\} \\
\text{CPD}_{\alpha,h}^{s,g} = \Big|\frac{1}{111\times (21-h)}\sum^{20}_{\omicron=h}\sum^{111}_{i=1}&\big[\mathds{1}\{d^{s,g}_{T_{\nu}+\omicron}(u_i) > d_{T_{\nu}+\omicron}^{s,g}(u_i)\} + \mathds{1}\{d^{s,g}_{T_{\nu}+\omicron}(u_i) < \widehat{d}^{s,g}_{T_{\nu}+\omicron}(u_i)\}\big]-\alpha\Big|.
\end{align*}

Both ECP and CPD are criteria for evaluating ECP, but they do not assess the sharpness of the prediction interval. By combining the coverage and sharpness, we consider a scoring rule for the prediction interval at age $u_i$, defined as
\begin{align*}
S_{\alpha} \big[\widehat{d}_{T_{\nu}+\omicron, \text{lb}}^{s,g}(u_i), \widehat{d}_{T_{\nu}+\omicron, \text{ub}}^{s,g}(u_i), d_{T_{\nu}+\omicron}^{s,g}(u_i)\big] =& \big[\widehat{d}_{T_{\nu}+\omicron, \text{ub}}^{s,g}(u_i) - \widehat{d}_{T_{\nu}+\omicron, \text{lb}}^{s,g}(u_i)\big] + \\
&\frac{2}{\alpha}\big[\widehat{d}_{T_{\nu}+\omicron, \text{lb}}^{s,g}(u_i) - d_{T_{\nu}+\omicron}^{s,g}(u_i)\big]\mathds{1}\big\{d_{T_{\nu}+\omicron}^{s,g}(u_i) < \widehat{d}_{T_{\nu}+\omicron, \text{lb}}^{s,g}(u_i)\big\} + \\
&\frac{2}{\alpha}\big[d_{T_{\nu}+\omicron}^{s,g}(u_i) - \widehat{d}_{T_{\nu}+\omicron, \text{ub}}^{s,g}(u_i)\big] \mathds{1}\big\{d_{T_{\nu}+\omicron}^{s,g}(u_i) > \widehat{d}_{T_{\nu}+\omicron, \text{ub}}^{s,g}(u_i)\big\}.
\end{align*}

\subsection{Interval forecast comparison}\label{sec:7.5}

\subsubsection{Empirical coverage probability}\label{sec:7.5.1}

\begin{figure}[!htb]
\centering
\includegraphics[width=8.71cm]{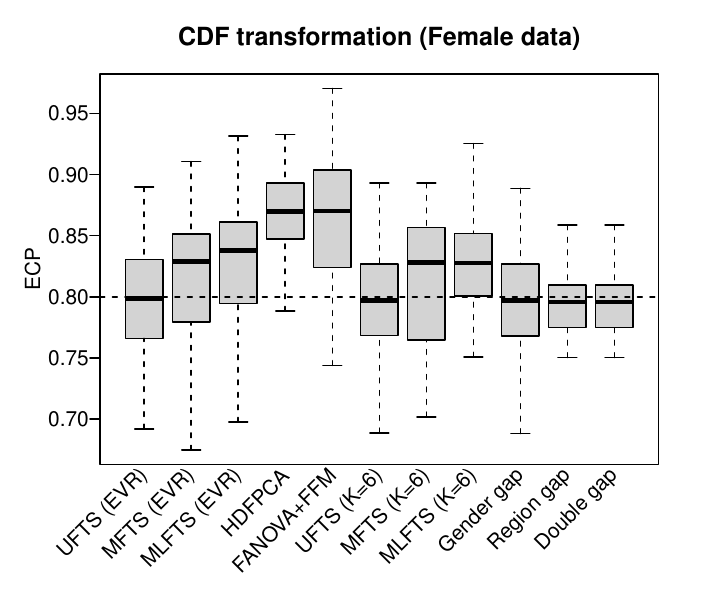}
\quad
\includegraphics[width=8.71cm]{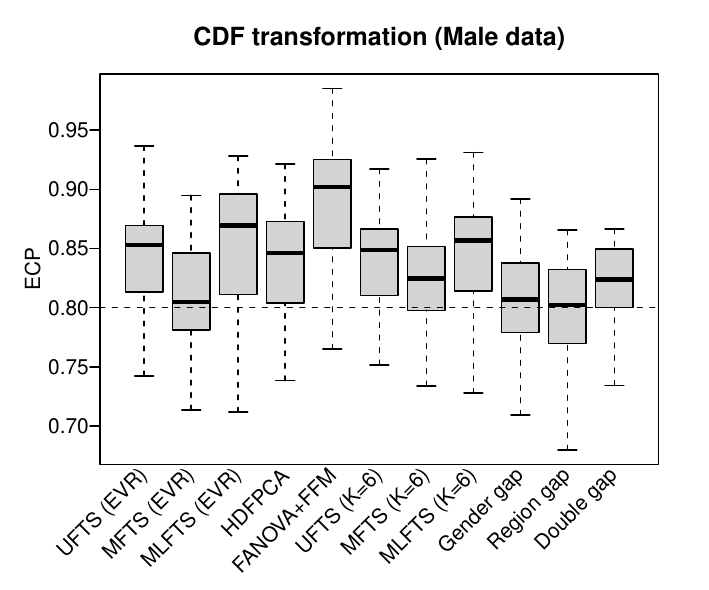}
\\
\includegraphics[width=8.71cm]{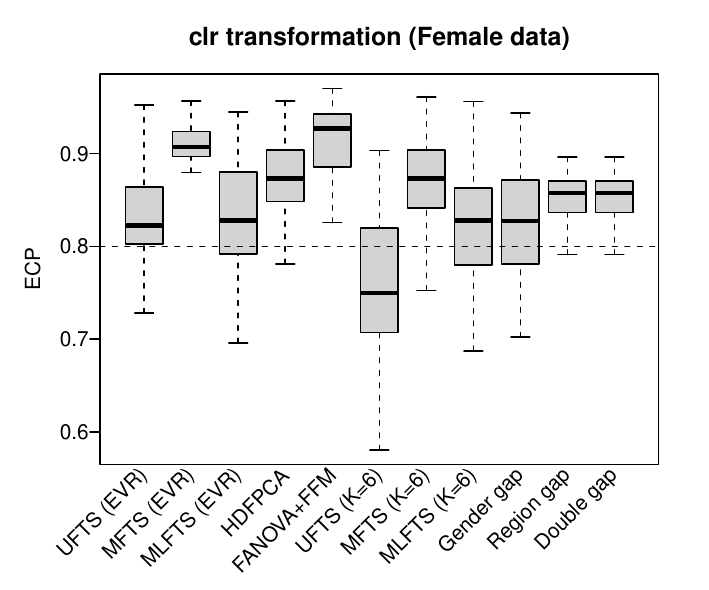}
\quad
\includegraphics[width=8.71cm]{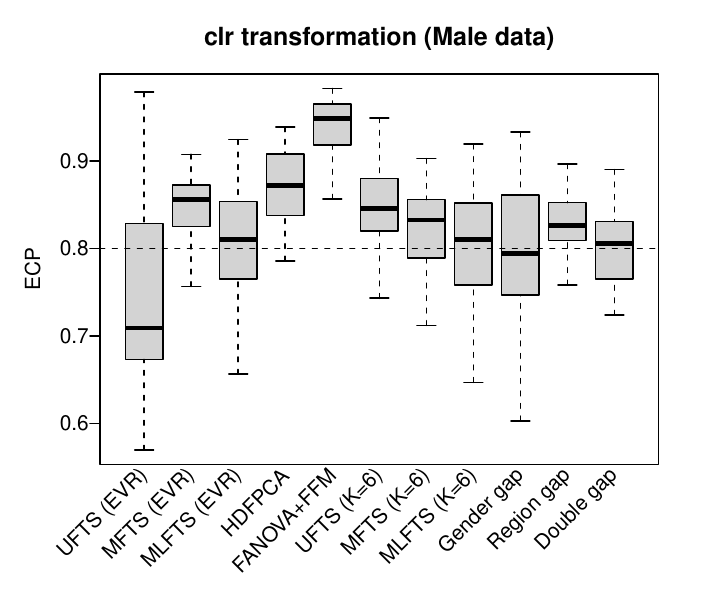}
\caption{\small At the 80\% nominal coverage probability, we compute the averaged empirical coverage probability, denoted by $\overline{\text{ECP}}_{\alpha=0.2}^{s,g}$, among the methods under the CDF and clr transformation for the Japanese female and male holdout test data.}\label{fig:ECP_boxplot}
\end{figure}

Averaging over the 20 forecast horizons, we compute the $\overline{\text{ECP}}^{s,g}_{\alpha}=\frac{1}{20}\sum^{20}_{h=1}\text{ECP}_{\alpha,h}^{s,g}$ for each gender at all prefectures. Using the resulting ECP values across the 47 prefectures, we construct the boxplots at the 80\% nominal coverage level, as shown in Figure~\ref{fig:ECP_boxplot}. 

For the female data, the region and double-gap methods are identical by construction and exhibit the smallest dispersion under the CDF transformation, with their medians lying at the nominal 80\% coverage level. For the male data, the MFTS (EVR) and region-gap methods are suggested under the CDF transformation.

\begin{figure}[!htb]
\centering
\includegraphics[width=8.71cm]{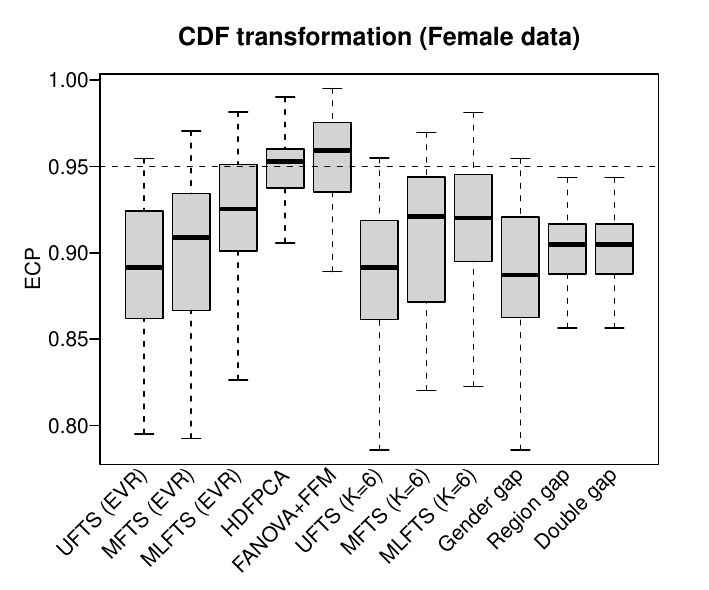}
\quad
\includegraphics[width=8.71cm]{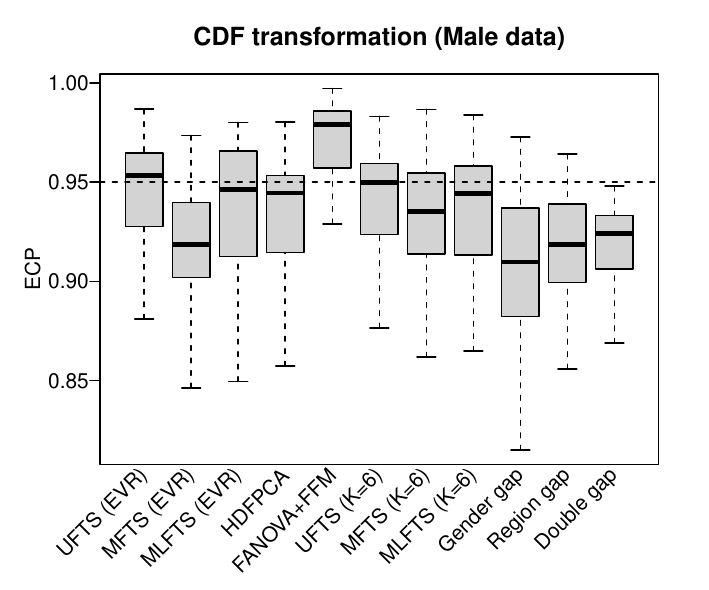}
\\
\includegraphics[width=8.71cm]{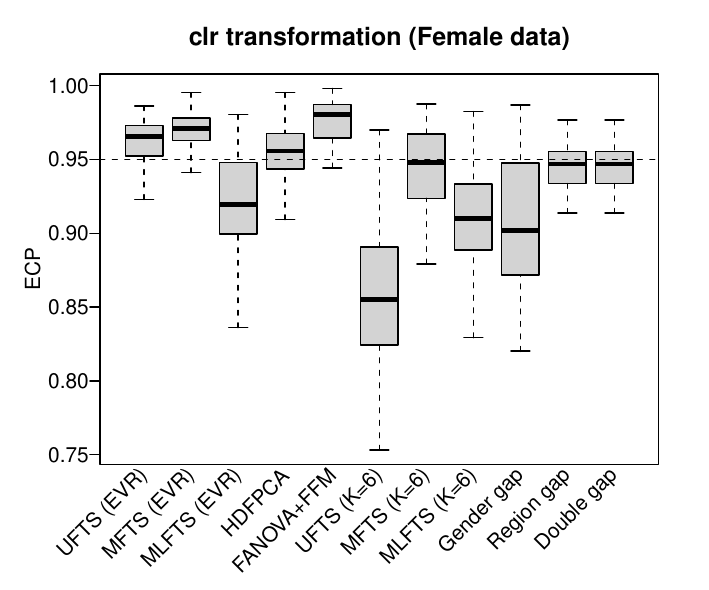}
\quad
\includegraphics[width=8.71cm]{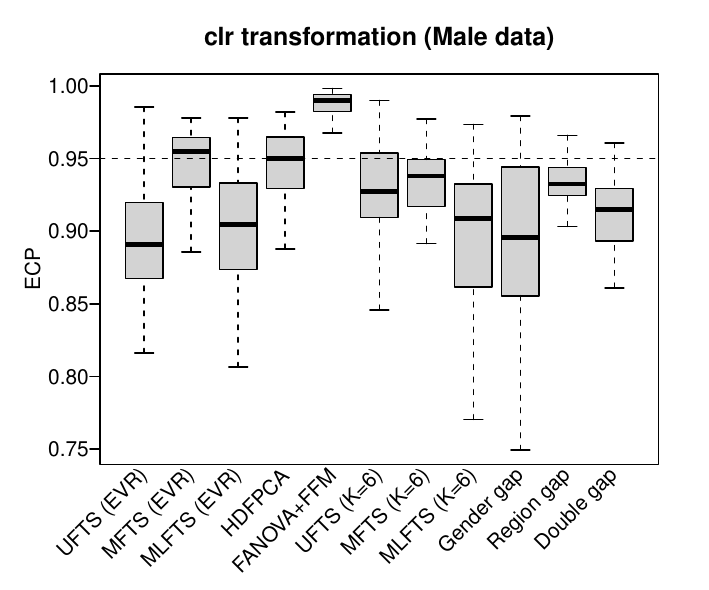}
\caption{\small At the 95\% nominal coverage probability, we compute the averaged empirical coverage probability, denoted by $\overline{\text{ECP}}_{\alpha=0.05}^{s,g}$, among the methods under the CDF and clr transformation for the Japanese female and male holdout test data.}\label{fig:ECP_95_boxplot}
\end{figure}

At the 95\% nominal coverage probability, the boxplots of the ECP values are presented in Figure~\ref{fig:ECP_95_boxplot}. For the female data, HDFPCA exhibits the best performance, achieving coverage close to the nominal 95\% level while also yielding the smallest dispersion under both transformations. Under the clr transformation, the region-gap and double-gap methods are also comparable. For the male data, the UFTS $(K = 6)$ is preferable under the CDF transformation, while the HDFPCA method is advocated under the clr transformation.

\subsubsection{Coverage probability difference}\label{sec:7.5.2}

Using the split conformal prediction method, we construct prediction interval for one-to-20-step ahead. Due to the limited space, we display selected forecast horizon, as well as two summary statistics across 20 horizons, at the nominal coverage probability of 80\% in Table~\ref{tab:4}. As measured by the CPD in Table~\ref{tab:4}, the gap modeling approach is generally preferred to multi-population approach. Between the CDF and clr transformations, the CDF transformation provides smaller CPD than the ones from the clr transformation.
\begin{small}
\tabcolsep 0.025in
\renewcommand{\arraystretch}{0.97}
\begin{longtable}{@{}lllrrrrrrrrrrr@{}}
\caption{\small Averaged over 47 prefectures, we evaluate and compare the interval forecast accuracy measured by the $\text{CPD}_{\alpha=0.2, h}^{s,g}$ under the CDF and clr transformations. For several methods, we compare the results under the EVR and $K=6$. The univariate time-series forecasting method is ETS. The method with the smallest overall error is bolded.}
\label{tab:4}\\
\toprule
& & & \multicolumn{3}{c}{EVR} & & FANOVA & \multicolumn{3}{c}{$K=6$} & \multicolumn{3}{c}{Gap} \\
\cmidrule(lr){4-6}\cmidrule(lr){9-11}\cmidrule(lr){12-14}
$\alpha$ & Sex & $h$ & UFTS & MFTS & MLFTS & HDFPCA & FFM & UFTS & MFTS & MLFTS & Gender & Region & Double \\
\midrule
\endfirsthead
\hline
$\alpha$ & Sex & $h$ & UFTS & MFTS & MLFTS & HDFPCA & FFM & UFTS & MFTS & MLFTS & Gender & Region & Double \\
\hline
\endhead
\hline
\multicolumn{14}{r}{{Continued on next page}} \\
\endfoot
\hline
\endlastfoot
\multicolumn{5}{l}{\hspace{-0.1in} \underline{CDF transformation (ETS)}} & \\
0.2 	& F 	& 1 & 0.0381 & 0.0620 & 0.0368 & 0.0435 & 0.0997 & 0.0376 & 0.0379 & 0.0400 & 0.0385 & 0.0406 & 0.0406 \\ 
	& 	& 5 & 0.0516 & 0.0469 & 0.0576 & 0.0643 & 0.0826 & 0.0533 & 0.0446 & 0.0545 & 0.0528 & 0.0316 & 0.0316 \\ 
	& 	& 10 & 0.0542 & 0.0499 & 0.0676 & 0.0968 & 0.0811 & 0.0558 & 0.0538 & 0.0602 & 0.0555 & 0.0328 & 0.0328 \\ 
	& 	& 15 & 0.0546 & 0.0782 & 0.0629 & 0.0710 & 0.0902 & 0.0534 & 0.0691 & 0.0603 & 0.0523 & 0.0422 & 0.0422 \\ 
	& 	& 20 & 0.0715 & 0.1044 & 0.0717 & 0.0957 & 0.0921 & 0.0662 & 0.0943 & 0.0740 & 0.0662 & 0.0503 & 0.0503 \\ 
\cmidrule{3-14}
	& 	& Mean & 0.0531 & 0.0652 & 0.0617 & 0.0811 & 0.0831 & 0.0534 & 0.0587 & 0.0589 & 0.0535 & \textBF{0.0365} & \textBF{0.0365} \\ 
	& 	& Median & 0.0532 & 0.0570 & 0.0646 & 0.0813 & 0.0855 & 0.0538 & 0.0545 & 0.0592 & 0.0537 & \textBF{0.0358} & \textBF{0.0358} \\ 
\midrule
	& M 	& 1 & 0.0453 & 0.0847 & 0.0410 & 0.0391 & 0.0980 & 0.0361 & 0.0403 & 0.0348 & 0.0395 & 0.0269 & 0.0330 \\ 
	& 	& 5 & 0.0507 & 0.0716 & 0.0479 & 0.0411 & 0.0791 & 0.0461 & 0.0482 & 0.0417 & 0.0364 & 0.0339 & 0.0331 \\ 
	& 	& 10 & 0.0588 & 0.0535 & 0.0823 & 0.0622 & 0.0923 & 0.0526 & 0.0473 & 0.0704 & 0.0401 & 0.0457 & 0.0435 \\ 
	& 	& 15 & 0.0822 & 0.0975 & 0.1187 & 0.0956 & 0.1454 & 0.0788 & 0.0926 & 0.1082 & 0.0449 & 0.0616 & 0.0452 \\ 
	& 	& 20 & 0.0979 & 0.1121 & 0.0920 & 0.0876 & 0.1228 & 0.0989 & 0.1100 & 0.0885 & 0.0611 & 0.0286 & 0.0465 \\ 
\cmidrule{3-14}
	& 	& Mean & 0.0698 & 0.0783 & 0.0817 & 0.0688 & 0.1096 & 0.0649 & 0.0668 & 0.0734 & 0.0431 & 0.0476 & \textBF{0.0422} \\ 
	& 	& Median & 0.0614 & 0.0743 & 0.0862 & 0.0670 & 0.0964 & 0.0543 & 0.0516 & 0.0736 & \textBF{0.0419} & 0.0460 & 0.0435 \\ 
\midrule
\multicolumn{5}{l}{\hspace{-0.1in} \underline{clr transformation (ETS)}} & \\
0.2 	& F 	& 1 & 0.0572 & 0.0446 & 0.0537 & 0.0548 & 0.1072 & 0.0855 & 0.0391 & 0.0361 & 0.0363 & 0.0426 & 0.0426 \\ 
	& 	& 5 & 0.0543 & 0.0683 & 0.0584 & 0.0631 & 0.1057 & 0.0832 & 0.0534 & 0.0527 & 0.0552 & 0.0362 & 0.0362 \\ 
	& 	& 10 & 0.0604 & 0.1241 & 0.0797 & 0.1032 & 0.1337 & 0.0731 & 0.0987 & 0.0691 & 0.0694 & 0.0744 & 0.0744 \\ 
	& 	& 15 & 0.0655 & 0.1427 & 0.0766 & 0.0868 & 0.1387 & 0.0683 & 0.1121 & 0.0620 & 0.0696 & 0.0882 & 0.0882 \\ 
	& 	& 20 & 0.0981 & 0.1247 & 0.1043 & 0.1135 & 0.0846 & 0.0882 & 0.1041 & 0.0895 & 0.0748 & 0.0654 & 0.0654 \\ 
\cmidrule{3-14}
	& 	&   Mean & \textBF{0.0615} & 0.1103 & 0.0742 & 0.0869 & 0.1187 & 0.0785 & 0.0848 & 0.0627 & 0.0637 & 0.0638 & 0.0638 \\ 
	& 	&   Median & 0.0604 & 0.1244 & 0.0768 & 0.0873 & 0.1243 & 0.0786 & 0.1001 & 0.0633 & 0.0685 & \textBF{0.0574} & \textBF{0.0574} \\ 
\midrule
	& M 	& 1 & 0.1555 & 0.0490 & 0.1076 & 0.0365 & 0.1036 & 0.0569 & 0.0513 & 0.0874 & 0.0914 & 0.0235 & 0.0954 \\ 
	& 	& 5 & 0.1441 & 0.0483 & 0.0698 & 0.0478 & 0.1094 & 0.0503 & 0.0485 & 0.0725 & 0.0635 & 0.0350 & 0.0733 \\ 
	& 	& 10 & 0.1402 & 0.0677 & 0.0618 & 0.0934 & 0.1527 & 0.0847 & 0.0586 & 0.0629 & 0.0779 & 0.0362 & 0.0774 \\ 
 	& 	& 15 & 0.1050 & 0.1117 & 0.0965 & 0.1280 & 0.1858 & 0.1097 & 0.0945 & 0.0922 & 0.0968 & 0.0942 & 0.1047 \\ 
 	& 	& 20 & 0.1085 & 0.1225 & 0.1144 & 0.1080 & 0.1355 & 0.1127 & 0.1086 & 0.1171 & 0.1032 & 0.0578 & 0.1045 \\ 
\cmidrule{3-14}
	& 	&   Mean & 0.1230 & 0.0833 & 0.0845 & 0.0881 & 0.1467 & 0.0825 & 0.0753 & 0.0851 & 0.0834 & \textBF{0.0585} & 0.0968 \\ 
	& 	&   Median & 0.1358 & 0.0751 & 0.0844 & 0.0977 & 0.1509 & 0.0892 & 0.0638 & 0.0823 & 0.0830 & \textBF{0.0426} & 0.0925 \\ 
\bottomrule
\end{longtable}
\end{small}

In Table~\ref{tab:5}, we present the CPD at some selected forecast horizons at the nominal coverage probability of 95\%. Between the two transformations, the clr transformation provides smaller CPD than the ones obtained from the CDF transformation for both females and males. The nominal coverage probability of 95\% has a smaller penalty of over-coverage than the nominal coverage probability of 80\%.
\begin{small}
\tabcolsep 0.025in
\renewcommand{\arraystretch}{1.02}
\begin{longtable}{@{}lllrrrrrrrrrrr@{}}
\caption{\small Averaged over 47 prefectures, we evaluate and compare the interval forecast accuracy measured by the $\text{CPD}_{\alpha=0.05,h}^{s,g}$ under the CDF and clr transformations. For several methods, we compare the results under the EVR and $K=6$. The univariate time-series forecasting method is ETS. The method with the smallest overall error is bolded.}
\label{tab:5}\\
\toprule
& & & \multicolumn{3}{c}{EVR} & & FANOVA & \multicolumn{3}{c}{$K=6$} & \multicolumn{3}{c}{Gap} \\
\cmidrule(lr){4-6}\cmidrule(lr){9-11}\cmidrule(lr){12-14}
$\alpha$ & Sex & $h$ & UFTS & MFTS & MLFTS & HDFPCA & FFM & UFTS & MFTS & MLFTS & Gender & Region & Double \\
\midrule
\endfirsthead
\hline
$\alpha$ & Sex & $h$ & UFTS & MFTS & MLFTS & HDFPCA & FFM & UFTS & MFTS & MLFTS & Gender & Region & Double \\
\hline
\endhead
\hline
\multicolumn{14}{r}{{Continued on next page}} \\
\endfoot
\hline
\endlastfoot
\multicolumn{5}{l}{\hspace{-0.1in} \underline{CDF transformation (ETS)}} & \\
0.05 & F 	& 1 & 0.0402 & 0.0836 & 0.0288 & 0.0399 & 0.0412 & 0.0354 & 0.0374 & 0.0302 & 0.0360 & 0.0188 & 0.0188 \\ 
	& 	& 5 & 0.0523 & 0.0693 & 0.0341 & 0.0290 & 0.0371 & 0.0520 & 0.0506 & 0.0395 & 0.0522 & 0.0409 & 0.0409 \\ 
	& 	& 10 & 0.0652 & 0.0523 & 0.0406 & 0.0325 & 0.0306 & 0.0644 & 0.0502 & 0.0437 & 0.0642 & 0.0386 & 0.0386 \\ 
	& 	& 15 & 0.0757 & 0.0476 & 0.0503 & 0.0340 & 0.0300 & 0.0760 & 0.0547 & 0.0534 & 0.0762 & 0.0704 & 0.0704 \\ 
	& 	& 20 & 0.0999 & 0.0436 & 0.0649 & 0.0335 & 0.0680 & 0.0978 & 0.0537 & 0.0673 & 0.0980 & 0.0508 & 0.0508 \\ 
\cmidrule{3-14}
	& 	& Mean & 0.0662 & 0.0580 & 0.0436 & \textBF{0.0335} & 0.0382 & 0.0656 & 0.0506 & 0.0473 & 0.0663 & 0.0513 & 0.0513 \\ 
	& 	& Median & 0.0666 & 0.0518 & 0.0422 & \textBF{0.0322} & 0.0378 & 0.0659 & 0.0507 & 0.0454 & 0.0659 & 0.0490 & 0.0490 \\ 
\midrule
	& M	& 1 & 0.0327 & 0.0914 & 0.0385 & 0.0401 & 0.0407 & 0.0288 & 0.0449 & 0.0333 & 0.0428 & 0.0265 & 0.0210 \\ 
	& 	& 5 & 0.0284 & 0.0761 & 0.0392 & 0.0378 & 0.0353 & 0.0279 & 0.0512 & 0.0395 & 0.0540 & 0.0395 & 0.0473 \\ 
	& 	& 10 & 0.0270 & 0.0411 & 0.0369 & 0.0307 & 0.0372 & 0.0274 & 0.0400 & 0.0350 & 0.0474 & 0.0550 & 0.0424 \\ 
	& 	& 15 & 0.0238 & 0.0341 & 0.0428 & 0.0274 & 0.0401 & 0.0225 & 0.0342 & 0.0375 & 0.0485 & 0.0496 & 0.0333 \\ 
	& 	& 20 & 0.0642 & 0.0371 & 0.0591 & 0.0538 & 0.0297 & 0.0632 & 0.0338 & 0.0642 & 0.0361 & 0.0290 & 0.0290 \\ 
\cmidrule{3-14}
	& 	& Mean & 0.0309 & 0.0526 & 0.0422 & 0.0355 & 0.0379 & \textBF{0.0305} & 0.0415 & 0.0405 & 0.0516 & 0.0438 & 0.0382 \\ 
	& 	& Median & \textBF{0.0285} & 0.0391 & 0.0393 & 0.0338 & 0.0365 & 0.0288 & 0.0390 & 0.0376 & 0.0512 & 0.0443 & 0.0426 \\ 
\midrule
\multicolumn{5}{l}{\hspace{-0.1in} \underline{clr transformation (ETS)}} & \\
0.05 & F 	&1 & 0.0253 & 0.0247 & 0.0420 & 0.0488 & 0.0346 & 0.0777 & 0.0350 & 0.0301 & 0.0260 & 0.0177 & 0.0177 \\ 
 	& 	& 5 & 0.0266 & 0.0238 & 0.0341 & 0.0297 & 0.0334 & 0.0932 & 0.0310 & 0.0347 & 0.0408 & 0.0242 & 0.0242 \\ 
	& 	& 10 & 0.0294 & 0.0297 & 0.0382 & 0.0349 & 0.0390 & 0.0958 & 0.0255 & 0.0380 & 0.0518 & 0.0163 & 0.0163 \\ 
 	& 	& 15 & 0.0290 & 0.0322 & 0.0527 & 0.0254 & 0.0339 & 0.0992 & 0.0294 & 0.0675 & 0.0686 & 0.0212 & 0.0212 \\ 
	& 	& 20 & 0.1829 & 0.0244 & 0.0944 & 0.0350 & 0.0629 & 0.1255 & 0.0495 & 0.1003 & 0.0919 & 0.0274 & 0.0274 \\ 
\cmidrule{3-14}
	& 	& Mean & 0.0396 & 0.0291 & 0.0457 & 0.0321 & 0.0370 & 0.0972 & 0.0335 & 0.0493 & 0.0560 & \textBF{0.0241} & \textBF{0.0241} \\ 
	& 	& Median & 0.0290 & 0.0288 & 0.0414 & 0.0315 & 0.0349 & 0.0962 & 0.0305 & 0.0392 & 0.0558 & \textBF{0.0214} & \textBF{0.0214} \\ 
\midrule
	& M 	&  1 & 0.0945 & 0.0450 & 0.0969 & 0.0354 & 0.0413 & 0.0552 & 0.0515 & 0.0788 & 0.0810 & 0.0231 & 0.0758 \\ 
	& 	& 5 & 0.0661 & 0.0426 & 0.0794 & 0.0349 & 0.0399 & 0.0471 & 0.0565 & 0.0750 & 0.0624 & 0.0301 & 0.0696 \\ 
 	& 	& 10 & 0.0523 & 0.0330 & 0.0492 & 0.0295 & 0.0451 & 0.0392 & 0.0326 & 0.0536 & 0.0595 & 0.0313 & 0.0768 \\ 
 	& 	& 15 & 0.0441 & 0.0329 & 0.0564 & 0.0327 & 0.0479 & 0.0395 & 0.0282 & 0.0647 & 0.0662 & 0.0195 & 0.0300 \\ 
 	& 	& 20 & 0.2129 & 0.0289 & 0.1037 & 0.0588 & 0.0282 & 0.0829 & 0.0320 & 0.1062 & 0.0866 & 0.0721 & 0.0258 \\ 
\cmidrule{3-14}
 	& 	&  Mean & 0.0669 & 0.0376 & 0.0686 & 0.0360 & 0.0428 & 0.0451 & 0.0418 & 0.0710 & 0.0686 & \textBF{0.0289} & 0.0585 \\ 
	& 	&   Median & 0.0582 & 0.0363 & 0.0619 & 0.0342 & 0.0427 & 0.0420 & 0.0348 & 0.0714 & 0.0656 & \textBF{0.0271} & 0.0641 \\ 
\end{longtable}
\end{small}

\subsubsection{Mean interval score}\label{sec:7.5.3}

In Table~\ref{tab:6}, we present the MIS values between the multi-population and gap modeling approaches under the CDF and clr transformations, at the nominal coverage probability of 80\%. Between the two transformations, the clr transformation generally provides the smaller MIS than the ones obtained from the CDF transformation.

\begin{small}
\tabcolsep 0.028in
\renewcommand{\arraystretch}{1}
\begin{longtable}{@{}lllrrrrrrrrrrr@{}}
\caption{\small Averaged over 47 prefectures, we evaluate and compare the interval forecast accuracy measured by the $\text{MIS}_{\alpha=0.2,h}^{s,g}$ under the CDF and clr transformations at the nominal coverage of 80\%. For several methods, we compare the results under the EVR and $K=6$. The univariate time-series forecasting method is ETS. The method with the smallest overall error is bolded.}\label{tab:6}\\
\toprule
& & & \multicolumn{3}{c}{EVR} & & FANOVA & \multicolumn{3}{c}{$K=6$} & \multicolumn{3}{c}{Gap} \\
\cmidrule(lr){4-6}\cmidrule(lr){9-11}\cmidrule(lr){12-14}
$\alpha$ & Sex & $h$ & UFTS & MFTS & MLFTS & HDFPCA & FFM & UFTS & MFTS & MLFTS & Gender & Region & Double \\
\midrule
\endfirsthead
\hline
$\alpha$ & Sex & $h$ & UFTS & MFTS & MLFTS & HDFPCA & FFM & UFTS & MFTS & MLFTS & Gender & Region & Double \\
\hline
\endhead
\hline
\multicolumn{14}{r}{{Continued on next page}} \\
\endfoot
\hline
\endlastfoot
\multicolumn{5}{l}{\hspace{-0.1in} \underline{CDF transformation (ETS)}} & \\
0.2 	& F 	& 1 & 240 & 405 & 235 & 237 & 250 & 225 & 243 & 225 & 225 & 171 & 171 \\ 
	& 	& 5 & 418 & 528 & 364 & 317 & 369 & 407 & 394 & 367 & 408 & 277 & 277 \\ 
	& 	& 10 & 767 & 665 & 614 & 503 & 565 & 751 & 606 & 628 & 751 & 431 & 431 \\ 
	& 	& 15 & 1415 & 920 & 1096 & 729 & 849 & 1395 & 964 & 1112 & 1393 & 739 & 739 \\ 
	& 	& 20 & 2986 & 1752 & 2549 & 1327 & 2168 & 2968 & 1943 & 2556 & 2958 & 1531 & 1531 \\ 
\cmidrule{3-14}
	& 	&  Mean & 1066 & 795 & 869 & \textBF{593} & 690 & 1051 & 762 & 874 & 1054 & 611 & 611 \\ 
	& 	&  Median & 816 & 688 & 653 & 502 & 588 & 799 & 637 & 669 & 801 & \textBF{450} & \textBF{450} \\ 
\midrule
	& M 	& 1 & 246 & 267 & 212 & 216 & 276 & 230 & 212 & 200 & 256 & 184 & 205 \\ 
	& 	& 5 & 341 & 358 & 295 & 316 & 365 & 335 & 319 & 299 & 505 & 291 & 409 \\ 
	& 	& 10 & 474 & 459 & 421 & 456 & 518 & 467 & 439 & 431 & 944 & 438 & 716 \\ 
	& 	& 15 & 637 & 590 & 608 & 633 & 748 & 632 & 588 & 606 & 1544 & 570 & 1049 \\ 
	& 	& 20 & 1397 & 1323 & 1486 & 1515 & 1962 & 1405 & 1370 & 1474 & 3097 & 1942 & 2167 \\ 
\cmidrule{3-14}
	& 	& Mean & 557 & 539 & 532 & 558 & 616 & 550 & \textBF{521} & 533 & 1177 & 543 & 848 \\ 
	& 	& Median & 488 & 473 & \textBF{437} & 472 & 533 & 481 & 453 & 446 & 997 & 455 & 749 \\ 
\midrule
\multicolumn{5}{l}{\hspace{-0.1in} \underline{clr transformation (ETS)}} & \\
0.2 	& F 	& 1 & 1041 & 262 & 319 & 248 & 246 & 411 & 216 & 267 & 224 & 192 & 192 \\ 
	& 	& 5 & 1250 & 341 & 466 & 317 & 349 & 720 & 324 & 426 & 391 & 287 & 287 \\ 
	& 	& 10 & 1564 & 431 & 669 & 567 & 556 & 1226 & 515 & 665 & 679 & 439 & 439 \\ 
	& 	& 15 & 1795 & 621 & 1090 & 693 & 844 & 2019 & 786 & 1148 & 1206 & 684 & 684 \\ 
	& 	& 20 & 2684 & 1273 & 2405 & 1328 & 1898 & 3639 & 2017 & 2554 & 2490 & 1283 & 1283 \\ 
\cmidrule{3-14}
	& 	&   Mean & 1616 & \textBF{523} & 884 & 607 & 647 & 1497 & 680 & 901 & 914 & 533 & 533 \\ 
	& 	&   Median & 1592 & \textBF{447} & 701 & 543 & 580 & 1288 & 549 & 699 & 725 & 455 & 455 \\ 
\midrule
	& M 	& 1 & 1026 & 236 & 306 & 213 & 262 & 303 & 214 & 293 & 289 & 201 & 273 \\ 
	& 	&   5 & 1175 & 329 & 412 & 326 & 324 & 432 & 327 & 451 & 382 & 299 & 364 \\ 
	& 	&  10 & 1347 & 424 & 556 & 465 & 528 & 572 & 430 & 591 & 545 & 409 & 470 \\ 
	& 	& 15 & 1469 & 558 & 922 & 655 & 833 & 825 & 567 & 964 & 836 & 553 & 602 \\ 
 	& 	& 20 & 2094 & 1257 & 2132 & 1458 & 1880 & 1270 & 1400 & 2226 & 1894 & 2183 & 1420 \\ 
\cmidrule{3-14}
	& 	&   Mean & 1377 & \textBF{496} & 759 & 560 & 655 & 635 & 517 & 815 & 707 & 564 & 548 \\ 
	& 	&   Median & 1364 & 436 & 582 & 484 & 547 & 577 & 447 & 617 & 569 & \textBF{420} & 475 \\ 
\bottomrule
\end{longtable}
\end{small}

In Table~\ref{tab:7}, we present the MIS for the multi-population and gap modeling approaches under the CDF and clr transformations at the nominal coverage probability of 95\%. Overall, the clr transformation yields smaller MIS values than the CDF transformation. For the female data, the region gap and double gap methods are recommended. For the male data, the region gap method outperforms the double gap method, although it is surpassed by the multi-population approach.

\begin{small}
\tabcolsep 0.027in
\renewcommand{\arraystretch}{0.97}
\begin{longtable}{@{}lllrrrrrrrrrrr@{}}
\caption{\small Averaged over 47 prefectures, we evaluate and compare the interval forecast accuracy measured by the $\text{MIS}_{\alpha=0.05,h}^{s,g}$ under the CDF and clr transformations at the nominal coverage probability of 95\%. For several methods, we compare the results under the EVR and $K=6$. The univariate time-series forecasting method is ETS. The method with the smallest overall error is bolded.}\label{tab:7}\\
\toprule
& & & \multicolumn{3}{c}{EVR} & & FANOVA & \multicolumn{3}{c}{$K=6$} & \multicolumn{3}{c}{Gap} \\
\cmidrule(lr){4-6}\cmidrule(lr){9-11}\cmidrule(lr){12-14}
$\alpha$ & Sex & $h$ & UFTS & MFTS & MLFTS & HDFPCA & FFM & UFTS & MFTS & MLFTS & Gender & Region & Double \\
\midrule
\endfirsthead
\hline
$\alpha$ & Sex & $h$ & UFTS & MFTS & MLFTS & HDFPCA & FFM & UFTS & MFTS & MLFTS & Gender & Region & Double \\
\hline
\endhead
\hline
\multicolumn{14}{r}{{Continued on next page}} \\
\endfoot
\hline
\endlastfoot
\multicolumn{5}{l}{\hspace{-0.1in} \underline{CDF transformation (ETS)}} & \\
0.05 & F 	&   1 & 402 & 859 & 372 & 399 & 398 & 379 & 427 & 371 & 380 & 276 & 276 \\ 
	& 	&   5 & 754 & 1094 & 589 & 505 & 586 & 731 & 718 & 614 & 735 & 423 & 423 \\ 
	& 	&   10 & 1552 & 1347 & 1090 & 826 & 941 & 1509 & 1186 & 1147 & 1508 & 674 & 674 \\ 
	& 	&   15 & 3238 & 1873 & 2223 & 1365 & 1377 & 3189 & 2006 & 2278 & 3170 & 1189 & 1189 \\ 
	& 	&   20 & 8004 & 4823 & 6768 & 3742 & 5694 & 7959 & 5326 & 7130 & 7916 & 3524 & 3524 \\ 
\cmidrule{3-14}
	& 	&   Mean & 2400 & 1703 & 1799 & 1122 & 1236 & 2364 & 1613 & 1839 & 2373 & \textBF{1079} & \textBF{1079} \\ 
	& 	&   Median & 1665 & 1403 & 1185 & 833 & 976 & 1620 & 1262 & 1244 & 1623 & \textBF{705} & \textBF{705} \\ 
\midrule
	& M 	&   1 & 363 & 481 & 328 & 331 & 430 & 328 & 341 & 311 & 415 & 271 & 311 \\ 
	& 	&   5 & 484 & 613 & 452 & 475 & 545 & 471 & 518 & 456 & 877 & 427 & 645 \\ 
	& 	&   10 & 699 & 758 & 670 & 710 & 836 & 683 & 710 & 683 & 1771 & 667 & 1125 \\ 
	& 	&   15 & 974 & 897 & 948 & 950 & 1295 & 965 & 900 & 930 & 3015 & 856 & 1575 \\ 
	& 	&   20 & 3150 & 3632 & 3396 & 3973 & 6420 & 3005 & 3758 & 3651 & 8776 & 5071 & 5165 \\ 
\cmidrule{3-14}
	& 	&   Mean & 930 & 975 & 924 & 978 & 1204 & \textBF{908} & 937 & 938 & 2457 & 958 & 1684 \\ 
	& 	&   Median & 720 & 778 & \textBF{694} & 739 & 855 & 706 & 730 & 706 & 1877 & 699 & 1188 \\ 
\midrule
\multicolumn{5}{l}{\hspace{-0.1in} \underline{clr transformation (ETS)}} & \\
 0.05 & F 	& 1 & 1312 & 390 & 528 & 418 & 395 & 834 & 369 & 444 & 369 & 291 & 291 \\ 
	& 	& 5 & 1739 & 528 & 846 & 508 & 575 & 1574 & 537 & 726 & 691 & 413 & 413 \\ 
	& 	& 10 & 2359 & 652 & 1213 & 959 & 930 & 2935 & 912 & 1167 & 1246 & 638 & 638 \\ 
 	& 	& 15 & 2744 & 928 & 2243 & 1301 & 1279 & 5130 & 1425 & 2370 & 2475 & 998 & 998 \\ 
 	& 	& 20 & 7874 & 3778 & 6765 & 4752 & 6832 & 10579 & 6065 & 7470 & 6723 & 3684 & 3684 \\ 
\cmidrule{3-14}
	& 	&   Mean & 2677 & 925 & 1839 & 1211 & 1244 & 3723 & 1404 & 1879 & 1899 & \textBF{902} & \textBF{902} \\ 
	& 	&  Median & 2411 & 672 & 1280 & 924 & 956 & 3104 & 995 & 1240 & 1354 & \textBF{667} & \textBF{667} \\ 
\midrule
	& M  &   1 & 1548 & 366 & 547 & 326 & 404 & 531 & 337 & 534 & 471 & 286 & 415 \\ 
	& 	&   5 & 1726 & 489 & 747 & 512 & 504 & 814 & 514 & 907 & 624 & 422 & 557 \\ 
	& 	&   10 & 1950 & 607 & 1010 & 708 & 866 & 1111 & 653 & 1128 & 920 & 602 & 693 \\ 
	& 	&   15 & 2143 & 819 & 1878 & 966 & 1358 & 1680 & 860 & 2001 & 1527 & 805 & 849 \\ 
	& 	&   20 & 5583 & 3221 & 5965 & 3981 & 4349 & 3564 & 3550 & 6051 & 4887 & 6047 & 3886 \\ 
\cmidrule{3-14}
	& 	&   Mean & 2205 & \textBF{821} & 1597 & 964 & 1139 & 1277 & 894 & 1785 & 1347 & 1013 & 929 \\ 
	& 	&   Median & 1994 & 625 & 1065 & 738 & 895 & 1099 & 688 & 1205 & 965 & \textBF{620} & 716 \\ 
\bottomrule
\end{longtable}
\end{small}

\subsection{Statistical significance of forecast accuracy}

\begin{figure}[!htb]
\centering
\includegraphics[width=\textwidth]{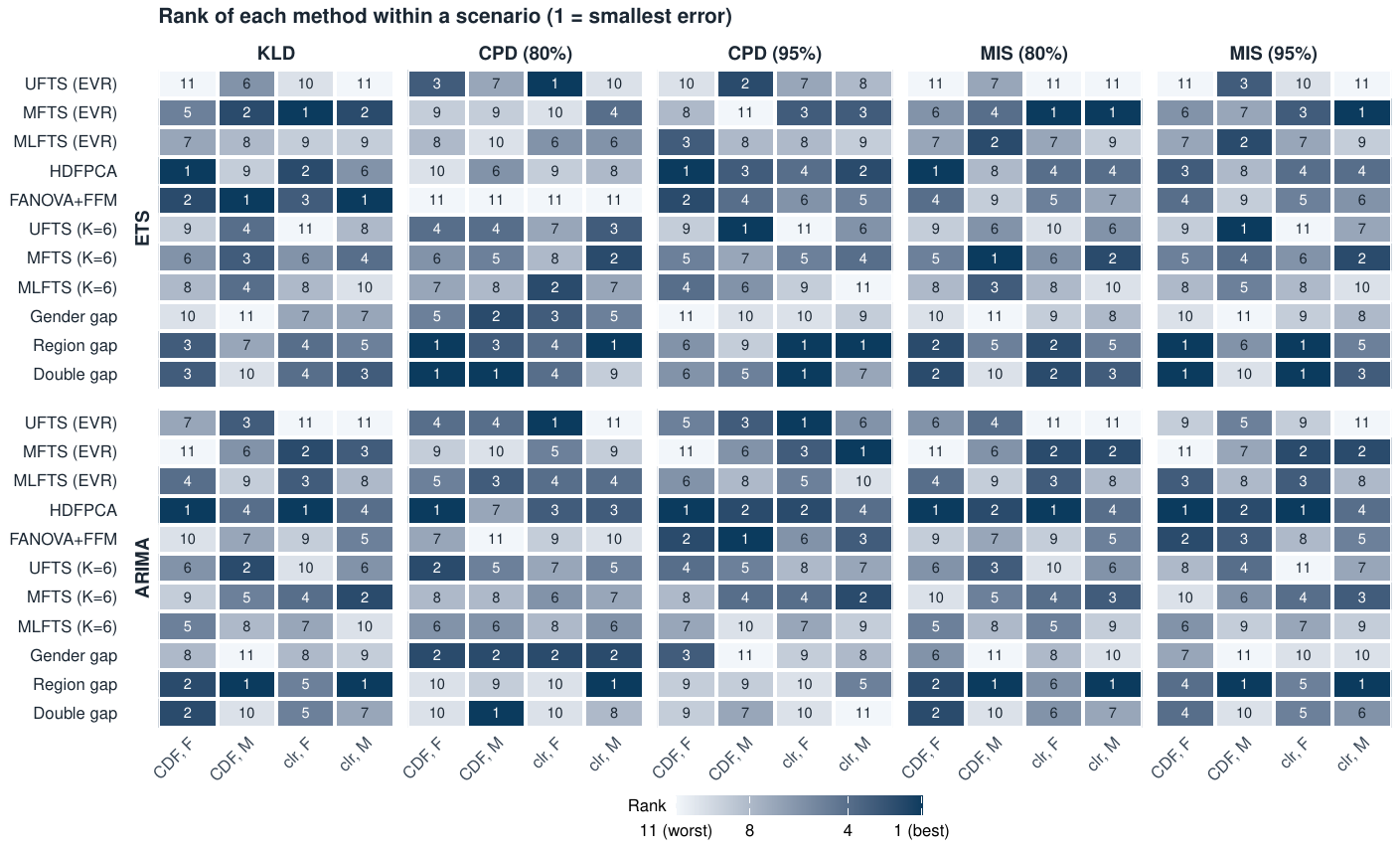}
\caption{\small Rank of each forecasting method within each of the 40 evaluation scenarios (five metrics $\times$ two transformations $\times$ two genders $\times$ two forecasting methods), with 1 denoting the smallest averaged error and darker cells better performance. Ties share the lower rank; the region-gap and double-gap methods coincide for the female data by construction.\label{fig:win_heatmap}}
\end{figure}

Figure~\ref{fig:win_heatmap} shows a win heat map of the point and interval forecast accuracy, where the 11 methods are ranked from best (1) to worst (11) within each of the 40 scenarios defined by metric, transformation, gender, and univariate forecasting method; the ranks are based on the mean errors, averaged over the 47 prefectures and the 20 forecast horizons. No method dominates: the best performer shifts with the choice of transformation, the evaluation metric, and the choice of univariate forecasting method. 

\begin{figure}[!htb]
\centering
\includegraphics[width=13.7cm]{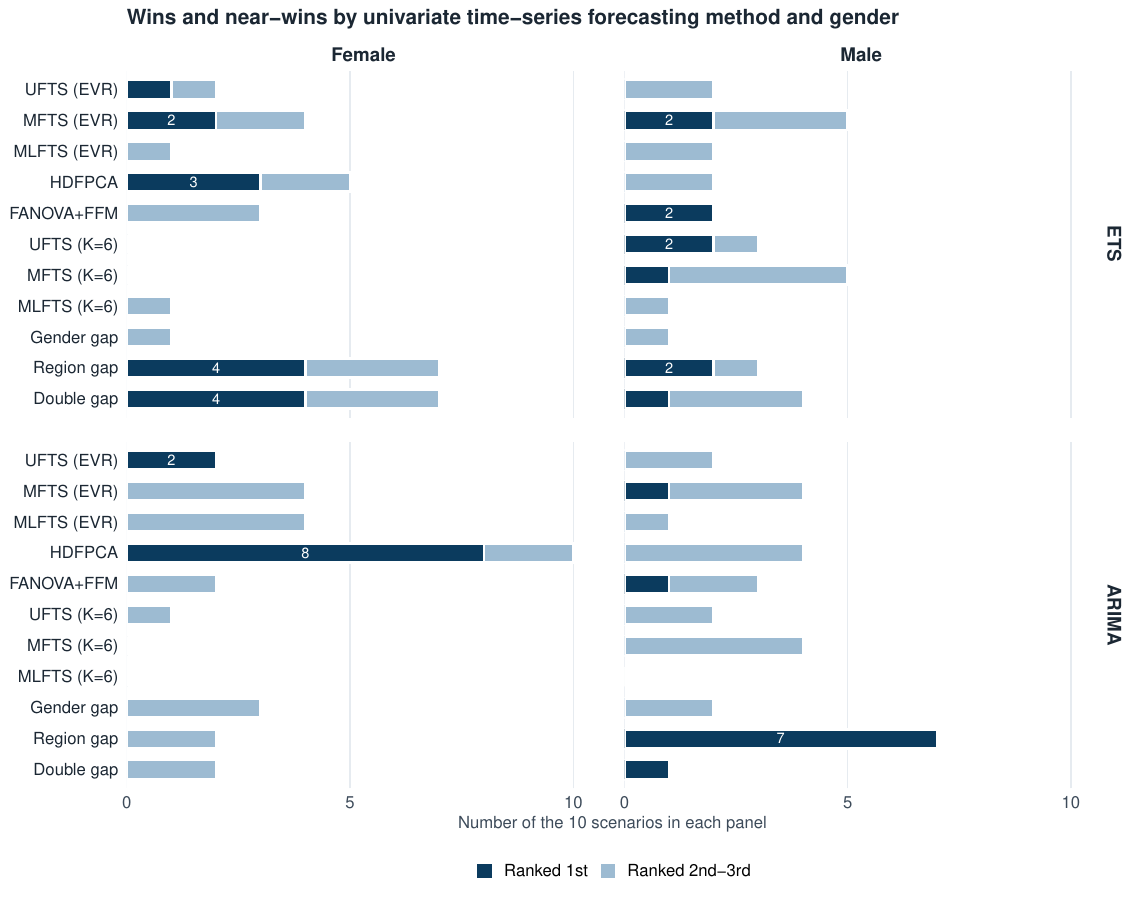}
\caption{\small  Number of scenarios in which each method ranks first, and second or third, by forecasting method and gender; each panel contains ten scenarios, formed by the five metrics and the two transformations.\label{fig:win_heatmap_count}}
\end{figure}

To summarize this pattern, Figure~\ref{fig:win_heatmap_count} counts how often each method ranks first, and second or third, separately for the ETS and ARIMA forecasts and for females and males. Also, it reveals that under ARIMA the wins collapse onto two methods: HDFPCA takes 8 of 10 female scenarios and region gap takes 7 of 10 male ones. Under ETS, the wins are spread across five distinct winners for females and six for males. The choice of univariate forecasting method not only shifts the ranking but also determines whether a clear winner exists.

We implement the model confidence set \citep{HLN11} to examine whether the differences in forecast accuracy are statistically significant. Computationally, we use the \textit{MCSprocedure} function in the MCS package in \Rlogo\ \citep{BC18}. In Table~\ref{tab:mcs}, we list the superior set of models at the 80\% confidence level using the $T_{\max}$ statistic with $B= 5{,}000$ bootstrap replications. 
\begin{table}[!htb]
\centering
\tabcolsep 0.15in
\renewcommand{\arraystretch}{0.8}
\begin{small}
\caption{\small Models retained in the model confidence set of \citet{HLN11} at the $80\%$ level, computed from the prefecture-level losses with
$B = 5{,}000$ bootstrap replications and the $T_{\max}$ statistic. A scenario is one combination of evaluation metric, nominal level, transformation, gender and univariate forecasting method. Note that region gap and double gap were deduplicated for the female data.} \label{tab:mcs}
\begin{tabular}{@{}ll L{2.3cm} L{2.8cm} L{2.7cm} L{2.6cm}@{}}
\toprule
& & \multicolumn{2}{c}{CDF transformation} & \multicolumn{2}{c}{clr transformation} \\
\cmidrule(lr){3-4}\cmidrule(lr){5-6}
Metric & Method & Female & Male & Female & Male \\
\midrule
\multirow{2}{*}{KLD}
 & ETS   & HDFPCA & FANOVA+FFM & MFTS (EVR) & FANOVA+FFM \\
 & ARIMA & HDFPCA & Region gap & HDFPCA & MFTS (EVR) MFTS ($K=6$) Region gap \\
\midrule
\multirow{2}{*}{CPD (80\%)}
 & ETS   & Region gap & Double gap & UFTS (EVR) MLFTS ($K=6$) Gender gap Region gap & Region gap \\
 & ARIMA & HDFPCA UFTS ($K=6$) Gender gap & Double gap & UFTS (EVR) & Region gap \\
\midrule
\multirow{2}{*}{CPD (95\%)}
 & ETS   & HDFPCA & UFTS ($K=6$) & Region gap & Region gap \\
 & ARIMA & HDFPCA & FANOVA+FFM & UFTS (EVR) & MFTS (EVR) \\
\midrule
\multirow{2}{*}{MIS (80\%)}
 & ETS   & HDFPCA & MFTS ($K=6$) & MFTS (EVR) & MFTS (EVR) \\
 & ARIMA & HDFPCA & Region gap & HDFPCA & Region gap \\
\midrule
\multirow{2}{*}{MIS (95\%)}
 & ETS   & Region gap & UFTS (EVR) UFTS ($K=6$) MFTS ($K=6$) MLFTS ($K=6$) Region gap & Region gap & MFTS (EVR) \\
 & ARIMA & HDFPCA & Region gap & HDFPCA & Region gap \\
\bottomrule
\end{tabular}
\end{small}
\end{table}

\section{Conclusion}\label{sec:8}

Using the age-specific Japanese subnational life-table death count, we compare the point and interval forecast accuracies under the CDF and clr transformations. While the CDF transformation can handle the presence of zero count, the clr transformation has theoretical foundation linking with Bayes Hilbert space. Both transformations are one-to-one mapping, and provide a tool for implementing algebraic operation in the transformed space. 

Within each transformation, we consider two approaches: the multi-population approach and the gap modeling approach. The multi-population approach accounts for potential correlations among the series, whereas the gap modeling approach focuses on the differences between factors, such as gender, region, and their interactions. The gap modeling approach also accommodates prefectures with different numbers of curves. To ensure a fair comparison between the two approaches, we use the same observation period, from 1973 to 2024.

In terms of point forecast accuracy, measured by the KLD, we find that the HDFPCA gives the smallest average KLD for the female data, while FANOVA+FFM gives the smallest average KLD for the male data. Under the clr transformation, MFTS (EVR) performs best for the female data, while FANOVA+FFM (or region/double gap) is strongest for the male data. Overall, the clr transformation generally outperforms the CDF transformation when paired with ETS forecasting method. Such superiority is horizon- and gender-dependent. As forecast horizon $h$ increases, KLD increases for all methods as expected from growing forecast uncertainty.  When ARIMA (Appendixes~\ref{sec:A} and~\ref{sec:B}) is used instead of ETS, HDFPCA is preferred for the female data, while the region gap approach is best for the male data under both transformations.

In terms of interval forecast accuracy, we consider three metrics, namely ECP, CPD, and MIS. At the 80\% nominal coverage probability, region/double gap methods under CDF transformation show the smallest dispersion and medians closest to nominal for the female data; MFTS (EVR) and region gap are best for the male data. At the 95\% nominal coverage probability, HDFPCA is best for the female data under both transformation, while UFTS $(K=6)$ under the CDF transformation and HDFPCA under the clr transformation are preferred for the male data. From the viewpoint of CPD, the gap modeling methods generally outperform multi-population methods. The CDF transformation gives smaller CPD at 80\% nominal coverage, but the clr transformation gives smaller CPD at the 95\% nominal coverage. Based on the MIS, the clr transformation generally yields smaller MIS than CDF transformation. At the 95\% nominal coverage probability, region and double gaps are recommended for the female data, while region gap outperforms double gap for the male data, though sometimes surpassed by multi-population approaches.

Between the two transformations, the clr transformation tends to perform better than the CDF transformation on most metrics, though the CDF transformation can still be competitive for CPD at lower nominal coverage probability. Between the two modeling approaches, gap modeling especially region gap is frequently the strongest performer for interval forecasts, while HDFPCA and FANOVA+FFM tend to lead on point forecast accuracy -- with the best method varying by gender, horizon and evaluation metric. At 80\% confidence levels, the MCS retains a single method in most scenarios, with HDFPCA and region gap the most frequent best performers. For reproducibility, the \Rlogo \ code is available at \url{https://github.com/hanshang/Age_distribution_death_modeling/}.

There are several ways in which the methodologies can be extended, and we briefly mention a few:
\begin{inparaenum}
\item[1)] We consider an extrinsic approach, based on transformations. Alternatively, one may explore some intrinsic approach, such as Wasserstein autoregressive model \citep{ZKP22}, Wasserstein exponential smoothing \citep{MJT+26}, and Wasserstein regression \citep{CLM23}. 
\item[2)] We model subnational life-table death count by age, gender and region. Alternatively, one may further explore other factors, such as socioeconomic status \citep{WCK21} and cause-of-death \citep{SM22}.
\end{inparaenum}

\section*{Acknowledgements}

The author gratefully acknowledges financial support from the Australian Research Council Future Fellowship (FT240100338) and the National Natural Science Foundation of China General Program (NSFC72673148).

\clearpage

\appendix

\section{Additional results on point forecast comparison (ARIMA)}\label{sec:A}

Using ARIMA as the time-series forecasting method, we evaluate and compare the point forecast accuracy, as measured by the KLD, among the multi-population and gap modeling approaches under the CDF and clr transformations. As shown in Table~\ref{tab:2}, for modeling the female data, the HDFPCA is suggested as it produces the smallest error, while the region gap modeling is recommended for the male data, under both transformations.
\begin{small}
\tabcolsep 0.038in
\renewcommand{\arraystretch}{0.95}
\begin{longtable}{@{}llrrrrrrrrrrr@{}}
\caption{\small Averaged over 47 prefectures, we evaluate and compare the point forecast accuracy measured by the KLD under the CDF and clr transformations. The univariate time-series forecasting method is ARIMA. The method with the smallest overall error is bolded. FANOVA+FFM represents the combination of two-way FANOVA and FFM.}\label{tab:2}\\
\toprule
& & \multicolumn{3}{c}{EVR} & & FANOVA & \multicolumn{3}{c}{$K=6$} & \multicolumn{3}{c}{Gap} \\
\cmidrule(lr){3-5}\cmidrule(lr){8-10}\cmidrule(lr){11-13}
Sex & $h$ & UFTS & MFTS & MLFTS & HDFPCA & FFM &
UFTS & MFTS & MLFTS & Gender & Region & Double \\
\midrule
\endfirsthead
\toprule
& & \multicolumn{3}{c}{EVR} & & FANOVA & \multicolumn{3}{c}{$K=6$} & \multicolumn{3}{c}{Gap} \\
\cmidrule(lr){3-5}\cmidrule(lr){8-10}\cmidrule(lr){11-13}
Sex & $h$ & UFTS & MFTS & MLFTS & HDFPCA & FFM &
UFTS & MFTS & MLFTS & Gender & Region & Double \\
\midrule
\endhead
\midrule
\multicolumn{13}{r}{Continued on next page}\\
\endfoot
\bottomrule
\endlastfoot
\multicolumn{5}{l}{\hspace{-0.1in} \underline{CDF transformation (ARIMA)}} & \\
F 	& 1 & 0.0059 & 0.0138 & 0.0059 & 0.0071 & 0.0053 & 0.0058 & 0.0106 & 0.0062 & 0.0062 & 0.0044 & 0.0044 \\ 
	& 5 & 0.0173 & 0.0350 & 0.0163 & 0.0124 & 0.0156 & 0.0172 & 0.0318 & 0.0183 & 0.0175 & 0.0119 & 0.0119 \\ 
	& 10 & 0.0540 & 0.0801 & 0.0495 & 0.0144 & 0.0515 & 0.0539 & 0.0795 & 0.0541 & 0.0544 & 0.0369 & 0.0369 \\ 
	& 15 & 0.1587 & 0.1603 & 0.1434 & 0.0455 & 0.1769 & 0.1586 & 0.1608 & 0.1491 & 0.1590 & 0.1392 & 0.1392 \\ 
	& 20 & 0.3903 & 0.3655 & 0.3573 & 0.0260 & 0.4387 & 0.3902 & 0.3662 & 0.3559 & 0.3902 & 0.3712 & 0.3712 \\ 
\cmidrule{2-13}
	& Mean & 0.1108 & 0.1190 & 0.1008 & \textBF{0.0243} & 0.1181 & 0.1107 & 0.1179 & 0.1043 & 0.1111 & 0.0971 & 0.0971 \\ 
	& Median & 0.0604 & 0.0868 & 0.0550 & \textBF{0.0170} & 0.0544 & 0.0603 & 0.0861 & 0.0599 & 0.0608 & 0.0415 & 0.0415 \\ 
\midrule
M 	& 1 & 0.0048 & 0.0060 & 0.0044 & 0.0043 & 0.0047 & 0.0047 & 0.0052 & 0.0043 & 0.0089 & 0.0037 & 0.0057 \\ 
	& 5 & 0.0084 & 0.0127 & 0.0092 & 0.0080 & 0.0094 & 0.0083 & 0.0117 & 0.0090 & 0.0365 & 0.0078 & 0.0315 \\ 
	& 10 & 0.0162 & 0.0269 & 0.0209 & 0.0160 & 0.0187 & 0.0161 & 0.0263 & 0.0202 & 0.1171 & 0.0169 & 0.1225 \\ 
	& 15 & 0.0416 & 0.0421 & 0.0570 & 0.0418 & 0.0455 & 0.0416 & 0.0422 & 0.0547 & 0.3164 & 0.0376 & 0.2719 \\ 
	& 20 & 0.1106 & 0.1056 & 0.1559 & 0.1159 & 0.1294 & 0.1106 & 0.1055 & 0.1495 & 0.8137 & 0.1081 & 0.7193 \\ 
\cmidrule{2-13}
	& Mean & 0.0315 & 0.0354 & 0.0428 & 0.0322 & 0.0355 & 0.0314 & 0.0349 & 0.0411 & 0.2247 & \textBF{0.0295} & 0.2019 \\ 
	& Median & 0.0175 & 0.0291 & 0.0228 & 0.0174 & 0.0192 & 0.0174 & 0.0284 & 0.0220 & 0.1305 & \textBF{0.0171} & 0.1372 \\ 
\midrule
\multicolumn{5}{l}{\hspace{-0.1in} \underline{clr transformation (ARIMA)}} & \\
F 	& 1 & 0.1041 & 0.0089 & 0.0255 & 0.0078 & 0.0051 & 0.0405 & 0.0081 & 0.0157 & 0.0062 & 0.0051 & 0.0051 \\ 
	& 5 & 0.1227 & 0.0292 & 0.0258 & 0.0112 & 0.0153 & 0.0417 & 0.0252 & 0.0207 & 0.0154 & 0.0112 & 0.0112 \\ 
	& 10 & 0.1522 & 0.0282 & 0.0335 & 0.0144 & 0.0529 & 0.0779 & 0.0573 & 0.0410 & 0.0442 & 0.0300 & 0.0300 \\ 
	& 15 & 0.1920 & 0.0562 & 0.0801 & 0.0362 & 0.1499 & 0.1651 & 0.0780 & 0.0979 & 0.1205 & 0.0968 & 0.0968 \\ 
	& 20 & 0.2273 & 0.1253 & 0.1739 & 0.0198 & 0.3356 & 0.3514 & 0.1995 & 0.2133 & 0.2786 & 0.2579 & 0.2579 \\ 
\cmidrule{2-13}
	& Mean & 0.1598 & 0.0491 & 0.0597 & \textBF{0.0201} & 0.0975 & 0.1229 & 0.0680 & 0.0706 & 0.0839 & 0.0687 & 0.0687 \\ 
	& Median & 0.1556 & 0.0437 & 0.0357 & \textBF{0.0144} & 0.0564 & 0.0848 & 0.0609 & 0.0447 & 0.0490 & 0.0331 & 0.0331 \\ 
\cmidrule{2-13}
M 	& 1 & 0.0551 & 0.0047 & 0.0118 & 0.0046 & 0.0041 & 0.0208 & 0.0074 & 0.0144 & 0.0066 & 0.0037 & 0.0069 \\ 
 	& 5 & 0.0678 & 0.0135 & 0.0157 & 0.0069 & 0.0066 & 0.0248 & 0.0117 & 0.0212 & 0.0107 & 0.0070 & 0.0122 \\ 
 	& 10 & 0.0882 & 0.0331 & 0.0257 & 0.0130 & 0.0159 & 0.0229 & 0.0285 & 0.0406 & 0.0244 & 0.0140 & 0.0238 \\ 
 	& 15 & 0.1108 & 0.0187 & 0.0591 & 0.0367 & 0.0401 & 0.0414 & 0.0211 & 0.1215 & 0.0691 & 0.0293 & 0.0523 \\ 
 	& 20 & 0.1353 & 0.0448 & 0.1436 & 0.1052 & 0.1249 & 0.0976 & 0.0555 & 0.1540 & 0.1846 & 0.0901 & 0.1660 \\ 
\cmidrule{2-13}
 	& Mean & 0.0911 & 0.0255 & 0.0455 & 0.0284 & 0.0319 & 0.0354 & 0.0248 & 0.0763 & 0.0512 & \textBF{0.0240} & 0.0440 \\ 
 	& Median & 0.0904 & 0.0233 & 0.0277 & 0.0141 & 0.0163 & 0.0239 & 0.0229 & 0.0447 & 0.0270 & \textBF{0.0140} & 0.0252 \\ 
\end{longtable}
\end{small}

\begin{figure}[!htb]
\centering
{\includegraphics[width=8.71cm]{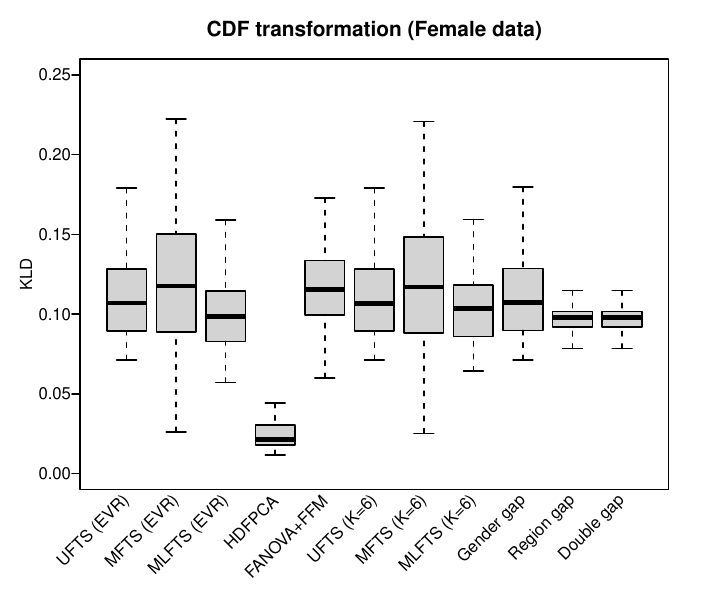}}
\quad
{\includegraphics[width=8.71cm]{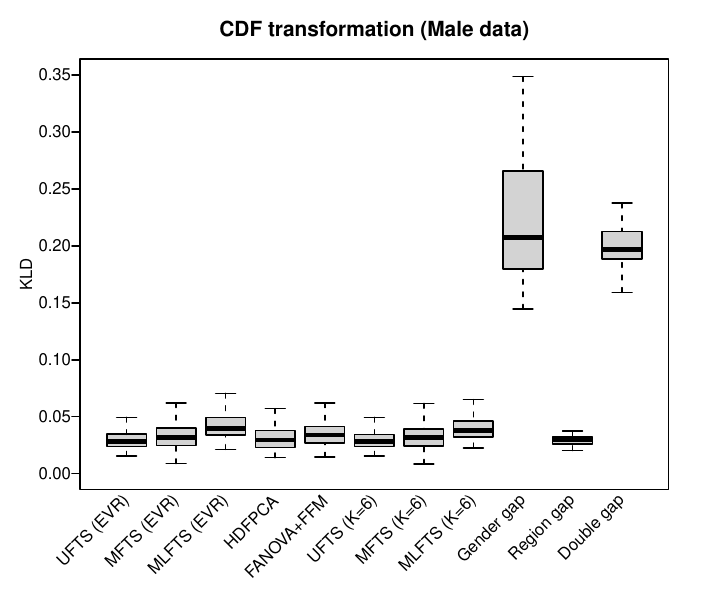}}
\\
{\includegraphics[width=8.71cm]{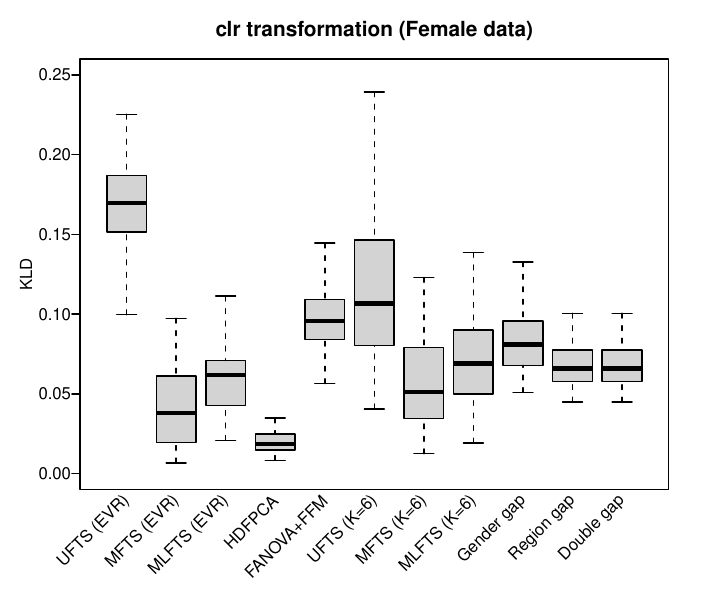}}
\quad
{\includegraphics[width=8.71cm]{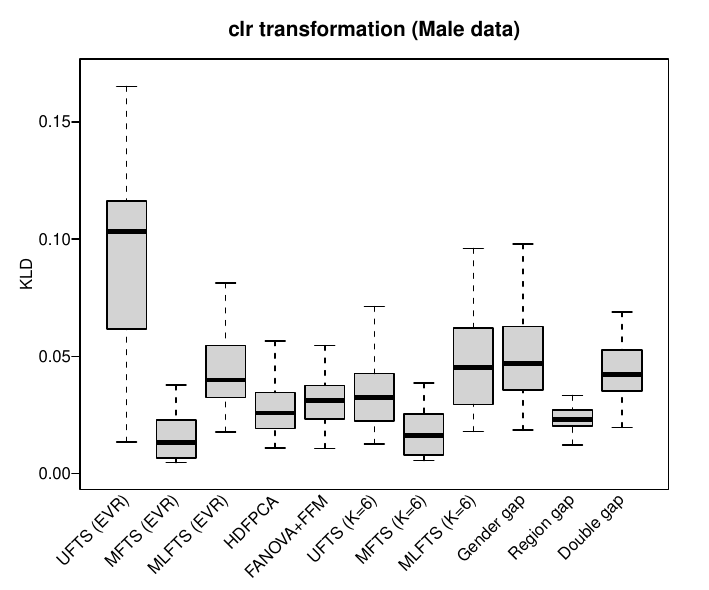}}
\caption{\small As measured by the $\overline{\text{KLD}}$, we compare the point forecast accuracy obtained from multi-population modeling and gap modeling approaches under the CDF and clr transformations.}\label{fig:4}
\end{figure}

In Figure~\ref{fig:4}, we present the $\overline{\text{KLD}}^{s,g}$ values for the female and male data, comparing the multi-population and gap modeling approaches under the CDF and clr transformations. For the male data, although the region gap modeling approach does not yield the smallest median $\overline{\text{KLD}}$, it produces the narrowest distribution. Nevertheless, when the results are summarized using either the mean or median, the region gap modeling approach achieves the smallest $\overline{\text{KLD}}$.

\clearpage

\section{Additional results on interval forecast accuracy (ARIMA)}\label{sec:B}

\subsection{Empirical coverage probability}\label{sec:B1}

\begin{figure}[!htb]
\centering
\includegraphics[width=8.71cm]{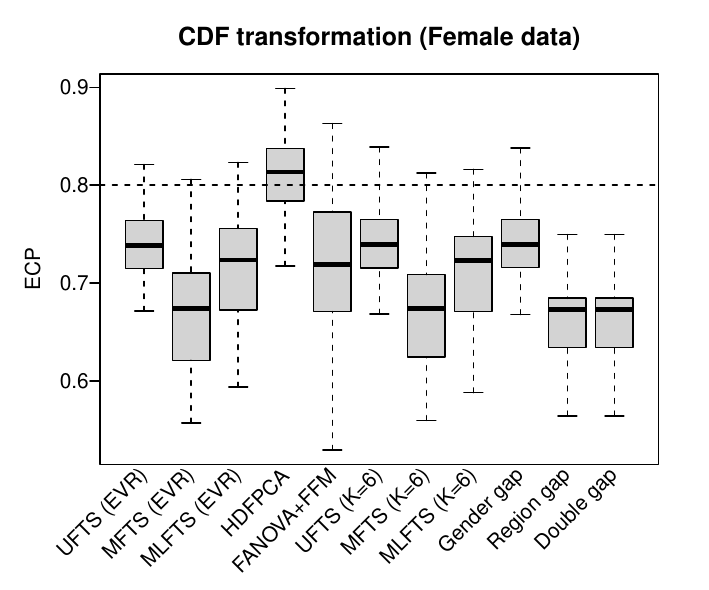}
\quad
\includegraphics[width=8.71cm]{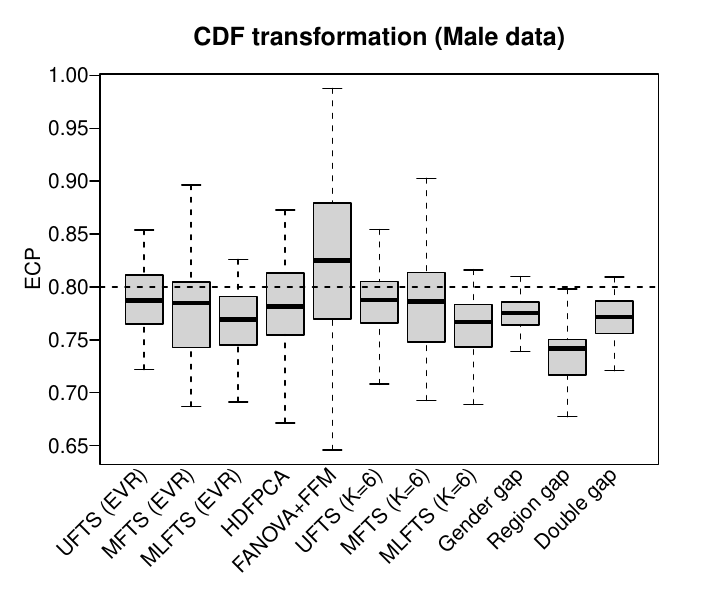}
\\
\includegraphics[width=8.71cm]{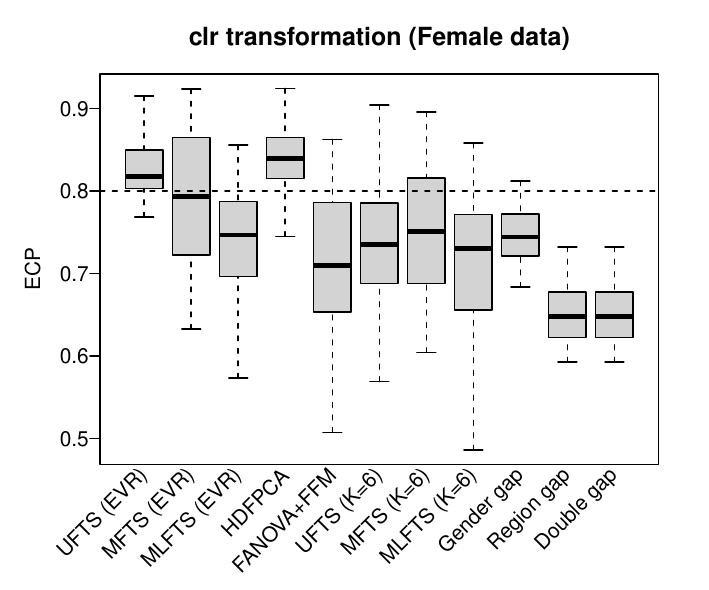}
\quad
\includegraphics[width=8.71cm]{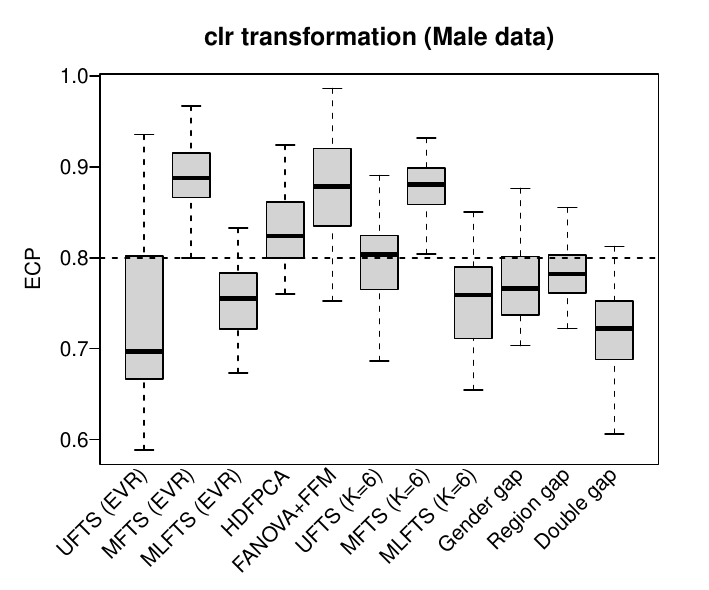}
\caption{\small At the 80\% nominal coverage probability, we compute the $\overline{\text{ECP}}^{s,g}_{\alpha=0.2}$ among the mulit-population and gap modeling approaches under the CDF and clr transformations for the Japanese subnational female and male holdout test data.}\label{fig:ARIMA_ECP_boxplot}
\end{figure}

In Figure~\ref{fig:ARIMA_ECP_boxplot}, we display the ECP values among the multi-population and gap modeling approaches under the two transformations. Compared with the ETS forecasting method, the ARIMA-based prediction intervals show weaker calibration: for the female data, most methods under the two transformations fall noticeably below the 80\% nominal coverage probability, with HDFPCA the closest to nominal and least variable. The pattern for male data is similiar but somewhat better calibrated, with gender gap and region gap closest to nominal level under the CDF transformation and clr transformation, respectively.

\begin{figure}[!htb]
\centering
\includegraphics[width=8.71cm]{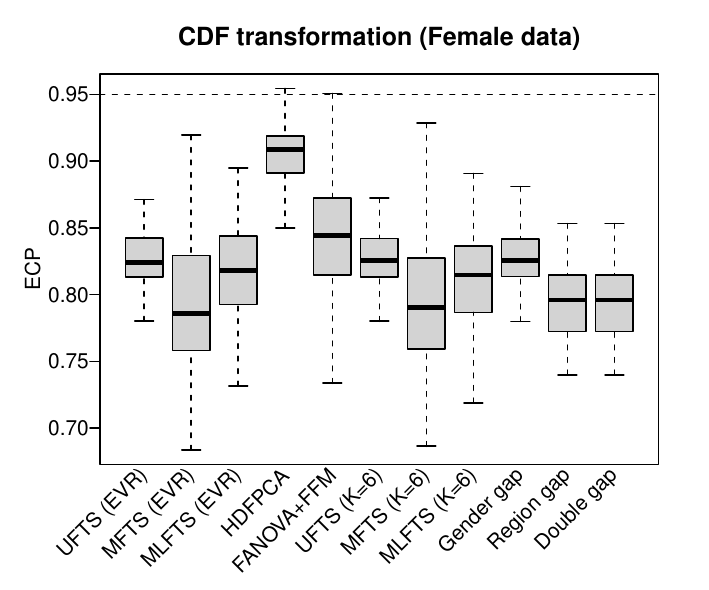}
\quad
\includegraphics[width=8.71cm]{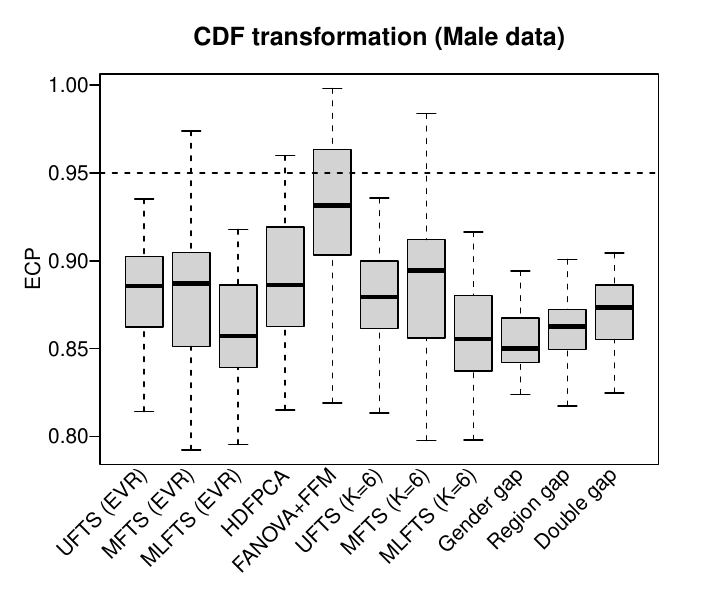}
\\
\includegraphics[width=8.71cm]{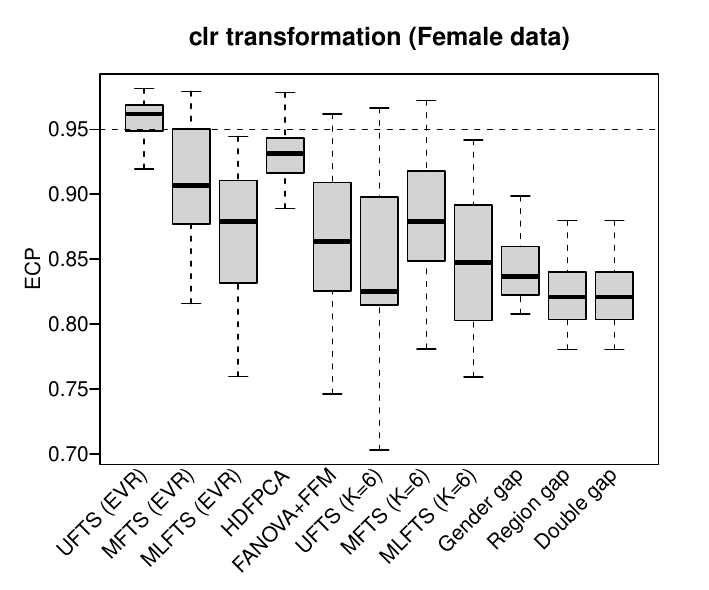}
\quad
\includegraphics[width=8.71cm]{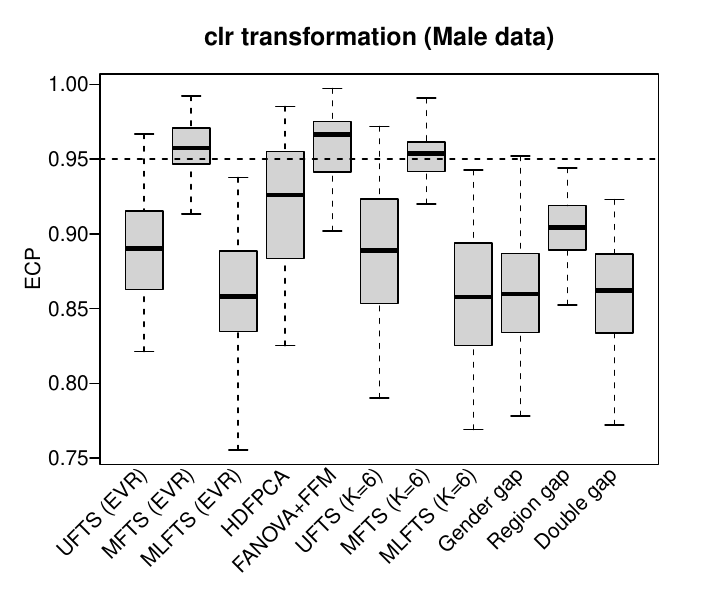}
\caption{\small At the 95\% nominal coverage probability, we compute the $\overline{\text{ECP}}_{\alpha=0.05}^{s,g}$ among the methods under the CDF and clr transformation for the Japanese subnational female and male holdout test data.}\label{fig:ARIMA_ECP_95_boxplot}
\end{figure}

In Figure~\ref{fig:ARIMA_ECP_95_boxplot}, we report the ECP at the 95\% nominal coverage probability for ARIMA-based forecasts, under the CDF and clr transformations for female and male data. As with the 80\% results in Figure~\ref{fig:ARIMA_ECP_boxplot}, ARIMA-based intervals are generally under-calibrated relative to the 95\% nominal level with wide dispersion across prefectures. For the female data, the HDFPCA stands out the most well-calibrated method, with median ECP closest to 95\% and the narrowest boxes. For the male data, FANOVA+FFM gives the median ECP closest to the 95\% nominal level, and MFTS (EVR or $K=6$) performs comparatively well under the clr transformation.

\subsection{Coverage probability difference}\label{sec:B2}

In Tables~\ref{tab:ARIMA_CPD_80} and~\ref{tab:ARIMA_CPD_95}, we report the CPD results under ARIMA forecasting methods, mirroring the ETS-based CPD analysis in the main text (Tables~\ref{tab:4} and~\ref{tab:5}) under the two transformation. At the 80\% nominal level in Table~\ref{tab:ARIMA_CPD_80}, no single method dominates. For the CDF transformation, HDFPCA gives the smallest averaged CPD for the female data while the gender and double gap models are best for the male data; under the clr transformation, UFTS (EVR) is best for the female data, while region and gender gap models perform best for the male data.
\begin{small}
\tabcolsep 0.028in
\renewcommand{\arraystretch}{1}
\begin{longtable}{@{}lllrrrrrrrrrrr@{}}
\caption{\small Averaged over 47 prefectures, we evaluate and compare the interval forecast accuracy measured by the $\overline{\text{CPD}}_{\alpha=0.2}^{s,g}$ under the CDF and clr transformations. The univariate time-series forecasting method is ARIMA. The method with the smallest overall error is bolded.}
\label{tab:ARIMA_CPD_80}\\
\toprule
& & & \multicolumn{3}{c}{EVR} & & FANOVA & \multicolumn{3}{c}{$K=6$} & \multicolumn{3}{c}{Gap} \\
\cmidrule(lr){4-6}\cmidrule(lr){9-11}\cmidrule(lr){12-14}
$\alpha$ & Sex & $h$ & UFTS & MFTS & MLFTS & HDFPCA & FFM & UFTS & MFTS & MLFTS & Gender & Region & Double \\
\midrule
\endfirsthead
\midrule
$\alpha$ & Sex & $h$ & UFTS & MFTS & MLFTS & HDFPCA & FFM & UFTS & MFTS & MLFTS & Gender & Region & Double \\
\midrule
\endhead
\midrule
\multicolumn{14}{r}{{Continued on next page}} \\
\endfoot
\bottomrule
\endlastfoot
\multicolumn{5}{l}{\hspace{-0.1in} \underline{CDF transformation (ARIMA)}} & \\
0.2 	& F 	& 1 & 0.0593 & 0.1139 & 0.0482 & 0.0736 & 0.0903 & 0.0547 & 0.0908 & 0.0573 & 0.0544 & 0.0286 & 0.0286 \\ 
	& 	& 5 & 0.0858 & 0.1464 & 0.1004 & 0.0702 & 0.0833 & 0.0858 & 0.1473 & 0.1093 & 0.0853 & 0.1074 & 0.1074 \\ 
	& 	& 10 & 0.0703 & 0.1232 & 0.0946 & 0.0905 & 0.0685 & 0.0705 & 0.1241 & 0.0962 & 0.0710 & 0.1204 & 0.1204 \\ 
	& 	& 15 & 0.0611 & 0.1299 & 0.0855 & 0.0522 & 0.1152 & 0.0612 & 0.1297 & 0.0857 & 0.0611 & 0.1591 & 0.1591 \\ 
	& 	& 20 & 0.0652 & 0.1468 & 0.1022 & 0.0662 & 0.1113 & 0.0652 & 0.1468 & 0.1000 & 0.0656 & 0.2246 & 0.2246 \\ 
	& 	& Mean & 0.0697 & 0.1361 & 0.0915 & \textBF{0.0680} & 0.1043 & 0.0694 & 0.1344 & 0.0958 & 0.0694 & 0.1365 & 0.1365 \\ 
	& 	& Median & 0.0671 & 0.1332 & 0.0926 & 0.0703 & 0.0897 & 0.0668 & 0.1337 & 0.0959 & \textBF{0.0666} & 0.1328 & 0.1328 \\ 
\midrule
	& M 	& 1 & 0.0504 & 0.0904 & 0.0352 & 0.0476 & 0.1005 & 0.0519 & 0.0599 & 0.0430 & 0.0388 & 0.0356 & 0.0283 \\ 
	& 	& 5 & 0.0501 & 0.0745 & 0.0416 & 0.0553 & 0.0694 & 0.0532 & 0.0689 & 0.0526 & 0.0490 & 0.0823 & 0.0477 \\ 
	& 	& 10 & 0.0360 & 0.0509 & 0.0316 & 0.0422 & 0.0633 & 0.0365 & 0.0525 & 0.0350 & 0.0311 & 0.0771 & 0.0255 \\ 
	& 	& 15 & 0.0419 & 0.0653 & 0.0338 & 0.0546 & 0.0901 & 0.0407 & 0.0655 & 0.0351 & 0.0205 & 0.0332 & 0.0248 \\ 
	& 	& 20 & 0.0495 & 0.0615 & 0.0710 & 0.0585 & 0.0789 & 0.0501 & 0.0617 & 0.0616 & 0.0433 & 0.0675 & 0.0517 \\ 
\cmidrule{3-14}
	& 	& Mean & 0.0446 & 0.0694 & 0.0425 & 0.0524 & 0.0797 & 0.0454 & 0.0656 & 0.0463 & 0.0348 & 0.0678 & \textBF{0.0338} \\ 
	& 	& Median & 0.0436 & 0.0697 & 0.0368 & 0.0538 & 0.0777 & 0.0427 & 0.0650 & 0.0443 & 0.0327 & 0.0672 & \textBF{0.0313} \\ 
\midrule
\multicolumn{5}{l}{\hspace{-0.1in} \underline{clr transformation (ARIMA)}} & \\
0.2 	& F 	& 1 & 0.0623 & 0.0527 & 0.0900 & 0.0593 & 0.0869 & 0.1166 & 0.0723 & 0.0886 & 0.0348 & 0.0451 & 0.0451 \\ 
  	& 	& 5 & 0.0472 & 0.0659 & 0.0732 & 0.0458 & 0.0908 & 0.0902 & 0.0809 & 0.1039 & 0.0582 & 0.0786 & 0.0786 \\ 
  	& 	& 10 & 0.0517 & 0.0840 & 0.0700 & 0.1266 & 0.0941 & 0.0810 & 0.0811 & 0.0957 & 0.0687 & 0.1220 & 0.1220 \\ 
  	& 	& 15 & 0.0504 & 0.1213 & 0.0930 & 0.0639 & 0.1653 & 0.0944 & 0.1038 & 0.1199 & 0.0855 & 0.2069 & 0.2069 \\ 
  	& 	& 20 & 0.1097 & 0.0981 & 0.1056 & 0.0886 & 0.1162 & 0.0887 & 0.1059 & 0.1134 & 0.0909 & 0.2299 & 0.2299 \\ 
\cmidrule{3-14}
  	& 	&   Mean & \textBF{0.0545} & 0.0906 & 0.0862 & 0.0817 & 0.1268 & 0.0918 & 0.0917 & 0.1077 & 0.0695 & 0.1481 & 0.1481 \\ 
	& 	&   Median & \textBF{0.0512} & 0.0875 & 0.0808 & 0.0699 & 0.1043 & 0.0915 & 0.0868 & 0.1032 & 0.0691 & 0.1339 & 0.1339 \\ 
\midrule
	& M & 1 & 0.1578 & 0.0411 & 0.1065 & 0.0371 & 0.1145 & 0.0997 & 0.0467 & 0.1059 & 0.0788 & 0.0323 & 0.1389 \\ 
 	& 	& 5 & 0.1442 & 0.0381 & 0.0766 & 0.0414 & 0.0966 & 0.0641 & 0.0329 & 0.0770 & 0.0431 & 0.0447 & 0.1164 \\ 
 	& 	& 10 & 0.1342 & 0.1060 & 0.0516 & 0.0601 & 0.1031 & 0.0605 & 0.0975 & 0.0542 & 0.0402 & 0.0402 & 0.0662 \\ 
 	& 	& 15 & 0.0862 & 0.1606 & 0.0520 & 0.0807 & 0.1231 & 0.0636 & 0.1492 & 0.0577 & 0.0510 & 0.0447 & 0.0551 \\ 
 	& 	& 20 & 0.1001 & 0.1038 & 0.0738 & 0.0668 & 0.1526 & 0.0690 & 0.0930 & 0.0774 & 0.0817 & 0.0504 & 0.1030 \\ 
\cmidrule{3-14}
  	& 	&   Mean & 0.1141 & 0.0993 & 0.0660 & 0.0593 & 0.1079 & 0.0662 & 0.0917 & 0.0690 & 0.0524 & \textBF{0.0455} & 0.0936 \\ 
  	& 	&   Median & 0.1314 & 0.1049 & 0.0574 & 0.0635 & 0.1054 & 0.0638 & 0.0953 & 0.0630 & 0.0469 & \textBF{0.0441} & 0.0906 \\ 
\end{longtable}
\end{small}

At the 95\% nominal level in Table~\ref{tab:ARIMA_CPD_95}, HDFPCA stands out for the female data under the CDF transformation, while FANOVA+FFM is best for male data; under the clr transformation, UFTS (EVR) is best for the female data and MFTS (EVR) is best for the male data. Compared with the ETS-based CPD results, we observe that gap modeling particularly region gap was generally the strongest performer.
\begin{small}
\tabcolsep 0.025in
\renewcommand{\arraystretch}{0.83}
\begin{longtable}{@{}lllrrrrrrrrrrr@{}}
\caption{\small Averaged over 47 prefectures, we evaluate and compare the interval forecast accuracy measured by the $\overline{\text{CPD}}_{\alpha=0.05}^{s,g}$ under the CDF and clr transformations. The univariate time-series forecasting method is ARIMA. The method with the smallest overall error is bolded.}
\label{tab:ARIMA_CPD_95}\\
\toprule
& & & \multicolumn{3}{c}{EVR} & & FANOVA & \multicolumn{3}{c}{$K=6$} & \multicolumn{3}{c}{Gap} \\
\cmidrule(lr){4-6}\cmidrule(lr){9-11}\cmidrule(lr){12-14}
$\alpha$ & Sex & $h$ & UFTS & MFTS & MLFTS & HDFPCA & FFM & UFTS & MFTS & MLFTS & Gender & Region & Double \\
\midrule
\endfirsthead
\midrule
$\alpha$ & Sex & $h$ & UFTS & MFTS & MLFTS & HDFPCA & FFM & UFTS & MFTS & MLFTS & Gender & Region & Double \\
\midrule
\endhead
\midrule
\multicolumn{14}{r}{{Continued on next page}} \\
\endfoot
\bottomrule
\endlastfoot
\multicolumn{5}{l}{\hspace{-0.1in} \underline{CDF transformation (ARIMA)}} & \\
0.05 & F 	&  1 & 0.0606 & 0.1266 & 0.0545 & 0.0832 & 0.0455 & 0.0576 & 0.1007 & 0.0683 & 0.0577 & 0.0304 & 0.0304 \\ 
	& 	&   5 & 0.1218 & 0.1669 & 0.1324 & 0.0941 & 0.0915 & 0.1218 & 0.1662 & 0.1404 & 0.1217 & 0.1239 & 0.1239 \\ 
	& 	&   10 & 0.1368 & 0.1636 & 0.1556 & 0.0301 & 0.1059 & 0.1370 & 0.1653 & 0.1556 & 0.1373 & 0.1589 & 0.1589 \\ 
	& 	&   15 & 0.1443 & 0.1652 & 0.1536 & 0.0975 & 0.1560 & 0.1443 & 0.1651 & 0.1579 & 0.1442 & 0.2126 & 0.2126 \\ 
	& 	&   20 & 0.1228 & 0.0940 & 0.1142 & 0.0361 & 0.0941 & 0.1230 & 0.0944 & 0.1126 & 0.1205 & 0.1120 & 0.1120 \\ 
\cmidrule{3-14}
	& 	&   Mean & 0.1247 & 0.1544 & 0.1351 & \textBF{0.0616} & 0.1162 & 0.1245 & 0.1523 & 0.1401 & 0.1244 & 0.1542 & 0.1542 \\ 
	& 	&   Median & 0.1336 & 0.1623 & 0.1482 & \textBF{0.0672} & 0.1105 & 0.1336 & 0.1636 & 0.1503 & 0.1337 & 0.1593 & 0.1593 \\ 
\midrule
	& M	& 1 & 0.0588 & 0.1056 & 0.0418 & 0.0546 & 0.0466 & 0.0583 & 0.0757 & 0.0518 & 0.0529 & 0.0429 & 0.0224 \\ 
	& 	&   5 & 0.0830 & 0.1142 & 0.0885 & 0.0832 & 0.0482 & 0.0860 & 0.1074 & 0.0986 & 0.1041 & 0.1043 & 0.0776 \\ 
	& 	&   10 & 0.0728 & 0.0841 & 0.0978 & 0.0676 & 0.0521 & 0.0748 & 0.0845 & 0.0996 & 0.1078 & 0.1183 & 0.0776 \\ 
	& 	&   15 & 0.0650 & 0.0507 & 0.0978 & 0.0521 & 0.0455 & 0.0666 & 0.0504 & 0.0961 & 0.0948 & 0.0888 & 0.1037 \\ 
	& 	&   20 & 0.0897 & 0.0478 & 0.1014 & 0.0752 & 0.0323 & 0.0886 & 0.0478 & 0.0922 & 0.0867 & 0.0246 & 0.0636 \\ 
\cmidrule{3-14}
	& 	&   Mean & 0.0726 & 0.0783 & 0.0909 & 0.0679 & \textBF{0.0530} & 0.0744 & 0.0739 & 0.0935 & 0.0953 & 0.0925 & 0.0813 \\ 
	& 	&   Median & 0.0721 & 0.0816 & 0.0965 & 0.0668 & \textBF{0.0514} & 0.0740 & 0.0775 & 0.0982 & 0.1017 & 0.0991 & 0.0850 \\ 
\midrule
\multicolumn{5}{l}{\hspace{-0.1in} \underline{clr transformation (ARIMA)}} & \\
0.05 & F 	&  1 & 0.0224 & 0.0444 & 0.0743 & 0.0635 & 0.0371 & 0.0994 & 0.0669 & 0.0750 & 0.0391 & 0.0416 & 0.0416 \\ 
 	& 	& 5 & 0.0251 & 0.0637 & 0.0746 & 0.0573 & 0.0742 & 0.1035 & 0.0837 & 0.0983 & 0.0919 & 0.0902 & 0.0902 \\ 
  	& 	& 10 & 0.0288 & 0.0545 & 0.0765 & 0.0377 & 0.0972 & 0.1075 & 0.0705 & 0.1021 & 0.1214 & 0.1343 & 0.1343 \\ 
  	& 	& 15 & 0.0264 & 0.0711 & 0.1041 & 0.0725 & 0.1269 & 0.1118 & 0.0831 & 0.1218 & 0.1405 & 0.1716 & 0.1716 \\ 
  	& 	& 20 & 0.1973 & 0.0524 & 0.1081 & 0.0240 & 0.0790 & 0.1112 & 0.0713 & 0.1110 & 0.1323 & 0.0999 & 0.0999 \\ 
\cmidrule{3-14}
	& 	&  Mean & \textBF{0.0406} & 0.0605 & 0.0845 & 0.0480 & 0.0999 & 0.1091 & 0.0772 & 0.1043 & 0.1110 & 0.1260 & 0.1260 \\ 
	& 	&  Median & \textBF{0.0282} & 0.0611 & 0.0778 & 0.0446 & 0.0926 & 0.1103 & 0.0758 & 0.1039 & 0.1198 & 0.1320 & 0.1320 \\ 
\midrule
  	& M 	& 1 & 0.0986 & 0.0394 & 0.0980 & 0.0430 & 0.0454 & 0.1014 & 0.0500 & 0.0982 & 0.0747 & 0.0363 & 0.1118 \\ 
  	& 	& 5 & 0.0667 & 0.0315 & 0.1152 & 0.0473 & 0.0335 & 0.0822 & 0.0385 & 0.1140 & 0.0746 & 0.0657 & 0.1199 \\ 
  	& 	& 10 & 0.0517 & 0.0323 & 0.0938 & 0.0426 & 0.0396 & 0.0621 & 0.0334 & 0.0934 & 0.0842 & 0.0641 & 0.0988 \\ 
  	& 	& 15 & 0.0417 & 0.0433 & 0.0758 & 0.0434 & 0.0421 & 0.0613 & 0.0402 & 0.0818 & 0.0978 & 0.0485 & 0.0736 \\ 
  	& 	& 20 & 0.1924 & 0.0421 & 0.1076 & 0.0876 & 0.0933 & 0.0890 & 0.0482 & 0.0969 & 0.1293 & 0.0208 & 0.0504 \\ 
\cmidrule{3-14}
  	& 	&   Mean & 0.0686 & \textBF{0.0367} & 0.0953 & 0.0488 & 0.0439 & 0.0729 & 0.0395 & 0.0952 & 0.0899 & 0.0546 & 0.0989 \\ 
  	& 	&   Median & 0.0593 & \textBF{0.0371} & 0.0959 & 0.0470 & 0.0408 & 0.0693 & 0.0387 & 0.0953 & 0.0870 & 0.0542 & 0.1047 \\	
\end{longtable}
\end{small}

\subsection{Mean interval score}\label{sec:B3}

We present the MIS under the ARIMA-based forecasting for the two transformation at the nominal coverage probabilities of 80\% and 95\% in Table~\ref{tab:ARIMA_MIS_80}. HDFPCA achieves the smallest average MIS for the female data across both transformations and both coverage probabilities, reflecting its narrower and better-calibrated intervals. For the male data, the region gap method produces the smallest average MIS across both transformations and coverage probabilities, though MFTS and HDFPCA are close competitors.
\begin{small}
\tabcolsep 0.026in
\renewcommand{\arraystretch}{0.93}
\begin{longtable}{@{}lllrrrrrrrrrrr@{}}
\caption{\small Averaged over 47 prefectures, we evaluate and compare the interval forecast accuracy measured by the $\overline{\text{MIS}}_{\alpha}^{s,g}$ under the CDF and clr transformations at the nominal coverage probabilities of 80\% and 95\%. The univariate time-series forecasting method is ARIMA. The method with the smallest overall error is bolded.}
\label{tab:ARIMA_MIS_80}\\
\toprule
& & & \multicolumn{3}{c}{EVR} & & FANOVA & \multicolumn{3}{c}{$K=6$} & \multicolumn{3}{c}{Gap} \\
\cmidrule(lr){4-6}\cmidrule(lr){9-11}\cmidrule(lr){12-14}
$\alpha$ & Sex & $h$ & UFTS & MFTS & MLFTS & HDFPCA & FFM & UFTS & MFTS & MLFTS & Gender & Region & Double \\
\midrule
\endfirsthead
\midrule
$\alpha$ & Sex & $h$ & UFTS & MFTS & MLFTS & HDFPCA & FFM & UFTS & MFTS & MLFTS & Gender & Region & Double \\
\midrule
\endhead
\midrule
\multicolumn{14}{r}{{Continued on next page}} \\
\endfoot
\bottomrule
\endlastfoot
\multicolumn{5}{l}{\hspace{-0.1in} \underline{CDF transformation (ARIMA)}} & \\
0.2 	& F 	& 1 & 272 & 546 & 275 & 327 & 263 & 266 & 426 & 290 & 267 & 185 & 185 \\ 
	& 	& 5 & 653 & 1010 & 643 & 500 & 590 & 651 & 958 & 699 & 651 & 490 & 490 \\ 
	& 	& 10 & 1320 & 1591 & 1300 & 471 & 1243 & 1319 & 1591 & 1361 & 1321 & 1086 & 1086 \\ 
	& 	& 15 & 2325 & 2365 & 2256 & 1207 & 2525 & 2325 & 2370 & 2300 & 2325 & 2414 & 2414 \\ 
	& 	& 20 & 3946 & 3337 & 3745 & 1257 & 3683 & 3943 & 3341 & 3742 & 3937 & 3743 & 3743 \\ 
\cmidrule{3-14}
	& 	& Mean & 1636 & 1775 & 1587 & \textBF{738} & 1662 & 1636 & 1749 & 1629 & 1636 & 1562 & 1562 \\ 
	& 	& Median & 1404 & 1664 & 1377 & \textBF{604} & 1278 & 1403 & 1662 & 1437 & 1405 & 1168 & 1168 \\ 
\midrule
	& M 	&  1 & 256 & 291 & 228 & 237 & 287 & 249 & 256 & 225 & 314 & 199 & 244 \\ 
	& 	&  5 & 370 & 460 & 387 & 360 & 434 & 367 & 441 & 381 & 769 & 364 & 677 \\ 
	& 	&  10 & 523 & 632 & 590 & 514 & 612 & 519 & 625 & 581 & 1538 & 567 & 1406 \\ 
	& 	&  15 & 846 & 826 & 973 & 809 & 863 & 846 & 827 & 958 & 2527 & 857 & 2399 \\ 
	& 	&  20 & 1921 & 1826 & 2132 & 1874 & 1799 & 1927 & 1826 & 2172 & 4586 & 1502 & 3586 \\ 
\cmidrule{3-14}
	& 	&   Mean & 705 & 727 & 781 & 676 & 733 & 701 & 716 & 773 & 1836 & \textBF{666} & 1662 \\ 
	& 	&   Median & 543 & 655 & 617 & \textBF{533} & 616 & 540 & 648 & 608 & 1632 & 561 & 1503 \\ 
\midrule
\multicolumn{5}{l}{\hspace{-0.1in} \underline{clr transformation (ARIMA)}} & \\	
0.2 	& F 	& 1 & 1085 & 304 & 479 & 283 & 253 & 770 & 298 & 445 & 259 & 229 & 229 \\ 
	& 	& 5 & 1268 & 526 & 565 & 409 & 504 & 890 & 560 & 608 & 536 & 424 & 424 \\ 
	& 	& 10 & 1566 & 745 & 761 & 511 & 1047 & 1383 & 852 & 908 & 1037 & 815 & 815 \\ 
	& 	& 15 & 1809 & 1065 & 1237 & 975 & 1995 & 2052 & 1311 & 1443 & 1785 & 1698 & 1698 \\ 
	& 	& 20 & 2925 & 1978 & 2428 & 1295 & 3119 & 3513 & 2460 & 2715 & 3316 & 2831 & 2831 \\ 
\cmidrule{3-14}
	& 	&  Mean & 1656 & 866 & 997 & \textBF{673} & 1332 & 1632 & 1022 & 1137 & 1303 & 1145 & 1145 \\ 
	& 	&   Median & 1590 & 770 & 793 & \textBF{553} & 1105 & 1447 & 886 & 952 & 1098 & 866 & 866 \\ 
\midrule
	& M  & 1 & 1038 & 221 & 409 & 227 & 268 & 521 & 222 & 419 & 301 & 207 & 334 \\ 
 	& 	& 5 & 1178 & 334 & 525 & 318 & 365 & 538 & 335 & 536 & 411 & 328 & 495 \\ 
  	& 	& 10 & 1360 & 478 & 656 & 456 & 559 & 603 & 481 & 684 & 652 & 470 & 679 \\ 
  	& 	& 15 & 1517 & 661 & 958 & 730 & 816 & 846 & 673 & 1038 & 1133 & 680 & 923 \\ 
  	& 	& 20 & 2319 & 1521 & 2048 & 1856 & 1725 & 1982 & 1594 & 2069 & 2530 & 1380 & 1884 \\ 
\cmidrule{3-14}
	& 	&  Mean & 1422 & 572 & 839 & 620 & 681 & 775 & 582 & 883 & 904 & \textBF{556} & 803 \\ 
	& 	&   Median & 1381 & 495 & 681 & 473 & 566 & 611 & 497 & 711 & 689 & \textBF{467} & 695 \\ 	
\midrule
\multicolumn{5}{l}{\hspace{-0.1in} \underline{CDF transformation (ARIMA)}} & \\
0.05 & F 	&   1 & 489 & 1354 & 479 & 626 & 435 & 475 & 968 & 531 & 479 & 300 & 300 \\ 
	& 	&   5 & 1476 & 2720 & 1415 & 1021 & 1152 & 1471 & 2525 & 1596 & 1472 & 1052 & 1052 \\ 
	& 	&  10 & 3276 & 4345 & 3251 & 810 & 2782 & 3275 & 4344 & 3450 & 3280 & 2824 & 2824 \\ 
	& 	&   15 & 5984 & 6270 & 5731 & 2721 & 6003 & 5988 & 6290 & 5901 & 5986 & 6799 & 6799 \\ 
	& 	&   20 & 10896 & 7252 & 9221 & 3551 & 8106 & 10897 & 7262 & 10021 & 10781 & 6416 & 6416 \\ 
\cmidrule{3-14}
	& 	&   Mean & 4108 & 4478 & 3815 & \textBF{1528} & 3716 & 4106 & 4386 & 3990 & 4101 & 3824 & 3824 \\ 
	& 	&   Median & 3516 & 4567 & 3464 & \textBF{1130} & 2921 & 3516 & 4558 & 3660 & 3521 & 3063 & 3063 \\ 
\midrule
	& M  &  1 & 415 & 570 & 362 & 373 & 463 & 402 & 476 & 365 & 501 & 307 & 366 \\ 
	& 	&  5 & 621 & 927 & 691 & 600 & 678 & 617 & 864 & 689 & 1492 & 620 & 1229 \\ 
	& 	&  10 & 832 & 1167 & 1062 & 813 & 912 & 827 & 1142 & 1053 & 3379 & 971 & 2904 \\ 
	& 	& 15 & 1441 & 1250 & 1707 & 1265 & 1227 & 1451 & 1252 & 1686 & 5340 & 1418 & 4983 \\ 
 	& 	& 20 & 4895 & 5430 & 5069 & 4914 & 4718 & 4955 & 5430 & 5374 & 13092 & 3298 & 7554 \\ 
\cmidrule{3-14}
	&	&   Mean & 1304 & 1383 & 1477 & 1218 & 1233 & 1302 & 1350 & 1479 & 4094 & \textBF{1123} & 3353 \\ 
	& 	&   Median & 860 & 1175 & 1109 & \textBF{839} & 914 & 855 & 1161 & 1098 & 3596 & 923 & 3132 \\ 
\midrule
\multicolumn{5}{l}{\hspace{-0.1in} \underline{clr transformation (ARIMA)}} & \\	
 0.05 & F 	& 1 & 1343 & 536 & 948 & 501 & 417 & 1729 & 559 & 871 & 461 & 370 & 370 \\ 
  	& 	& 5 & 1762 & 990 & 1107 & 723 & 839 & 1992 & 1055 & 1203 & 1056 & 764 & 764 \\ 
  	& 	& 10 & 2356 & 1365 & 1357 & 917 & 1868 & 3247 & 1638 & 1763 & 2216 & 1584 & 1584 \\ 
  	& 	& 15 & 2744 & 1902 & 2275 & 1855 & 3784 & 5012 & 2570 & 2838 & 4004 & 3773 & 3773 \\ 
  	& 	& 20 & 8790 & 5202 & 6119 & 4666 & 8274 & 9534 & 6343 & 7323 & 9397 & 5512 & 5512 \\ 
\cmidrule{3-14}
  	& 	& Mean & 2781 & 1649 & 1920 & \textBF{1353} & 2580 & 3957 & 2054 & 2336 & 2918 & 2265 & 2265 \\ 
 	& 	&   Median & 2405 & 1415 & 1407 & \textBF{941} & 1919 & 3429 & 1718 & 1854 & 2360 & 1677 & 1677 \\ 
\midrule
  	& M 	& 1 & 1571 & 371 & 797 & 355 & 427 & 1211 & 396 & 850 & 495 & 317 & 582 \\ 
  	& 	& 5 & 1733 & 570 & 1040 & 501 & 559 & 1120 & 557 & 1099 & 689 & 558 & 945 \\ 
  	& 	& 10 & 1976 & 774 & 1166 & 669 & 815 & 1042 & 772 & 1287 & 1153 & 769 & 1240 \\ 
  	& 	& 15 & 2187 & 963 & 1584 & 1058 & 1181 & 1406 & 974 & 1848 & 2110 & 999 & 1454 \\ 
  	& 	& 20 & 6245 & 3613 & 5776 & 4428 & 3532 & 5821 & 3779 & 5290 & 6466 & 3667 & 4546 \\ 
\cmidrule{3-14}
  	& 	& Mean & 2308 & 996 & 1628 & 1047 & 1072 & 1535 & 1008 & 1767 & 1779 & \textBF{936} & 1429 \\ 
	& 	&  Median & 2027 & 800 & 1211 & \textBF{691} & 819 & 1112 & 795 & 1335 & 1230 & 734 & 1261 \\ 
\end{longtable}
\end{small}

\clearpage
\bibliographystyle{agsm}
\bibliography{ADD_modeling}

\end{document}